\documentclass[12pt]{article}

\usepackage{epsf}

\newcommand{\bmat}{\left(\begin{array}}
\newcommand{\emat}{\end{array}\right)}
\def\NPB{Nucl. Phys. B}
\def\PLB{Phys. Lett. B}
\def\PLB{Phys. Lett. B}

\def\yzero{\smash{\hbox{$y\kern-4pt\raise1pt\hbox{${}^\circ$}$}}}

\def\a{\alpha}
\def\b{\beta}
\def\l{\label}

\def\beq{\begin{equation}}
\def\eeq{\end{equation}}
\def\beqa{\begin{eqnarray}}
\def\eeqa{\end{eqnarray}}
\def\t{\times}

\def\th{\theta}

\def\-{\hphantom{-}}
\def\ov{\overline}
\def\s2{\frac{1}{\sqrt2}}

\def\beq{\begin{equation}}
\def\eeq{\end{equation}}
\def\beqa{\begin{eqnarray}}
\def\eeqa{\end{eqnarray}}

\def\IF{\relax{\rm I\kern-.18em F}}
\def\II{\relax{\rm I\kern-.18em I}}
\def\IP{\relax{\rm I\kern-.18em P}}
\def\IC{\relax\hbox{\kern.25em$\inbar\kern-.3em{\rm C}$}}
\def\IR{\relax{\rm I\kern-.18em R}}

\def\cp{{\cal P}}

\def\ch{{\cal H}}

\def\Dsl{\,\raise.15ex\hbox{/}\mkern-13.5mu D} 
\def\IZ{Z\kern-.4em  Z}

 \def\cp#1{\relax\ifmmode {\IP\kern-2pt{}_{#1}}\else $\IP\kern-2pt{}_{#1}$\=fi}

\catcode`\@=11
\newdimen\@rotdimen
\newbox\@rotbox

\def\@vspec#1{\special{ps:#1}}
\def\@rotstart#1{\@vspec{gsave currentpoint currentpoint translate
   #1 neg exch neg exch translate}}
\def\@rotfinish{\@vspec{currentpoint grestore moveto}}
\def\@rotr#1{\@rotdimen=\ht#1\advance\@rotdimen by\dp#1%
   \hbox to\@rotdimen{\hskip\ht#1\vbox to\wd#1{\@rotstart{90 rotate}%
   \box#1\vss}\hss}\@rotfinish}
\def\@rotl#1{\@rotdimen=\ht#1\advance\@rotdimen by\dp#1%
   \hbox to\@rotdimen{\vbox to\wd#1{\vskip\wd#1\@rotstart{270 rotate}%
   \box#1\vss}\hss}\@rotfinish}%
\def\@rotu#1{\@rotdimen=\ht#1\advance\@rotdimen by\dp#1%
   \hbox to\wd#1{\hskip\wd#1\vbox to\@rotdimen{\vskip\@rotdimen
   \@rotstart{-1 dup scale}\box#1\vss}\hss}\@rotfinish}%
\def\@rotf#1{\hbox to\wd#1{\hskip\wd#1\@rotstart{-1 1 scale}%
   \box#1\hss}\@rotfinish}%
\def\rotate{\@ifnextchar[{\@rotate}{\@rotate[l]}}
\def\@rotate[#1]#2{\setbox\@rotbox=\hbox{#2}\@nameuse{@rot#1}\@rotbox}

\catcode`\@=12

\begin{document}

\makeatletter \@addtoreset{equation}{section} \makeatother
\renewcommand{\theequation}{\thesection.\arabic{equation}}
\pagestyle{empty}
\pagestyle{empty}
\vspace{0.5in}
\rightline{FTUAM-02/23}
\rightline{IFT-UAM/CSIC-02-39}
\rightline{\today}
\vspace{2.0cm}
\setcounter{footnote}{0}

\begin{center}
\LARGE{
{\bf Deformed Intersecting D6-Brane GUTS I}}
\\[4mm]
{\large{ Christos ~Kokorelis
}
\\[1mm]}
\normalsize{\em Departamento de F\'\i sica Te\'orica C-XI and 
Instituto de F\'\i sica 
Te\'orica C-XVI,}
,\\[-0.3em]
{\em Universidad Aut\'onoma de Madrid, Cantoblanco, 28049, Madrid, Spain}
\end{center}
\vspace{1.0mm}


\begin{center}
{\small  ABSTRACT}
\end{center}
By employing D6-branes
intersecting at angles in $D = 4$ type IIA strings, 
we construct {\em four stack string GUT
models} (PS-I class), that contain 
at low energy {\em exactly the three generation Standard model}
 with no extra matter and/or
extra gauge group factors.
These classes of models are based on the Pati-Salam (PS)
gauge group
$SU(4)_C \times SU(2)_L \times SU(2)_R$.
They represent deformations around the quark and lepton
basic intersection number structure.
The models possess the same phenomenological characteristics
of some recently discussed examples (PS-A class) of
four stack
PS GUTS. Namely,
there are no colour triplet couplings to mediate proton decay
 and proton is stable
 as baryon number is a gauged symmetry. Neutrinos get masses
 of the correct sizes. Also the mass relation
 $m_e = m_d$ at the GUT scale is recovered.
 
 Moreover, we clarify the novel role of {\em extra}
 branes, the latter
 having non-trivial intersection numbers with
 quarks and leptons and creating scalar singlets, needed
 for
 the satisfaction of RR tadpole cancellation conditions.
 The presence of N=1 supersymmetry in sectors involving
 the {\em extra} branes is equivalent to the, model
 dependent, orthogonality conditions of the U(1)'s
 surviving massless the generalized Green-Schwarz
 mechanism.
 The use of
 {\em extra} branes
 creates mass couplings that predict the appearance
 of light fermion doublets up to the scale of
 electroweak scale symmetry breaking.

\newpage
\setcounter{page}{1} \pagestyle{plain}
\renewcommand{\thefootnote}{\arabic{footnote}}
\setcounter{footnote}{0}

\section{Introduction}

Major problems of string 
theory include among others the hierarchy of scale and 
particle masses after 
supersymmetry breaking. 
These phenomenological issues have by far been explored in the context of 
construction of semirealistic supersymmetric models 
of 
weakly coupled $N=1$ (orbifold) compactifications of the 
heterotic string theories \cite{ena}. In these theories 
one of the unsolved problems was the fact that 
the string scale which is of the order of $10^{18}$ GeV
was in clear disagreement with the observed unification of
gauge
coupling constants in the MSSM of $10^{16}$ GeV.
The latter problem remains a mystery 
even though
the observed discrepancy between the two high 
scales was attributed \footnote{among other options,} to
the presence of the $N=1$
string threshold corrections
to the gauge coupling constants \cite{dikomou}.

On the contrary in type I models, the string scale, which 
is a free parameter, can be lowered in the
TeV range \cite{antoba} thus suggesting 
that non-SUSY models with a string scale in the TeV region
is a viable possibility.
In this spirit, recently some new constructions have
appeared in a type I
string vacuum background which use 
intersecting branes \cite{tessera} and give four
dimensional non-supersymmetric models.

In these open string models \cite{tessera}
the use of background fluxes
in a D9 brane type I background \footnote{
In the T-dual 
language these backgrounds are represented by
D6 branes wrapping 3-cycles on a dual torus
and intersecting each other at certain angles. }.
breaks supersymmetry on the 
brane and gives chiral fermions 
with an even number of 
generations \cite{tessera}.
The fermions on those models get localized
in the
intersections between
branes \cite{tessera1}, \cite{bele}.
The introduction of a quantized background NS-NS B 
field \cite{eksi1,eksi2,eksi3}, that makes the tori
tilted
gives rise to 
semirealistic models with 
three generations \cite{tessera2}.
It should be noted that these backgrounds
are T-dual to models with magnetic
deformations \cite{carlo}.
Additional non-SUSY constructions in the context 
of intersecting branes,
from IIB orientifolds, consisting of getting at 
low energy the standard model 
spectrum with extra matter and additional chiral fermions
were derived in \cite{luis1}.
Also SUSY constructions
in the context of intersecting branes were considered in
\cite{uran}. In addition, 
constructions involving intersecting branes in compact Calabi-Yau spaces
were discussed in \cite{bra1}, while intersecting brane constructions 
in the context
of non-compact Calabi-Yau spaces were considered in \cite{angel}.
For some other work in the context of intersecting
branes see  \cite{cim, alda, alda1, kranio, maria, chi}. 
For some recent attempts to construct \footnote{non-SUSY GUTS in the context 
of type 
IIB with branes on singularities see \cite{yanagida}.}, non-SUSY GUT models in the
context of
intersecting
branes see \cite{blume, king, nano}.

Furthermore, an important step was taken in \cite{louis2},
by showing how to construct
the standard model (SM) spectrum together with right handed 
neutrinos in a systematic way. The authors considered,
as a
starting point,
IIA theory compactified on $T^6$ \cite{tessera}
assigned with an
orientifold
product $\Omega \times R$, where $\Omega$ is the worldsheet parity
operator
and R is the reflection operator with respect to one of the axis of
each tori.
In this case, the four stack D6-branes contain Minkowski space and  
each of the 
three remaining dimensions is wrapped up on a different $T^2$ torus.
In this
construction the proton is stable since the baryon number
is a gauged
$U(1)$ global symmetry.
A special feature of these models is that the neutrinos
can only get Dirac mass. These models have been
generalized to classes of models with
just the SM at low energy and having
five stacks \cite{kokos} and six stacks \cite{kokos2} of
 D6-branes at the string scale.
The models of \cite{kokos}, \cite{kokos2} are build as
deformations of the QCD intersection numbers, namely
they are build around the left and right handed quarks
intersection numbers. Also, they hold exactly the same
phenomenological properties of \cite{louis2}. Also,
these models -
due to
the presence of $N=1$ supersymmetric sectors necessary for
the breaking of the $U(1)$'s surviving massless the
Green-Schwarz mechanism - predict the unique
existence of one
supersymmetric partner of the right neutrino
and two supersymmetric partners of the right neutrino in the
 five and six stack SM's of \cite{kokos}, \cite{kokos2}
 respectively.

In addition, in \cite{kokos1} we presented the first
examples of classes
of string derived GUT models (PS-A class)
 that break completely to the SM at
low energies.
The models are developed in the same D6-brane
backgrounds as the SM's of \cite{louis2}. They
are based in the Pati-Salam (PS) \cite{Pati-Salam} GUT
structure $G_{422}$, 
$SU(4)_c \times SU(2)_L \times SU(2)_R$.
The models predict uniquely the existence
of light weak fermion doublets with energy between the
range 90 - 246 GeV, that is they can be directly tested at
present or future accelerators.
We also note another recent construction with D5 branes
intersecting at angles on an orientifold of
 $T^2 \times T^4/Z_N$ \cite{crema}. In this case, four stack
 models of D5 branes give just the SM at low
 energy. A full study of the latter models including
 an extension
 to five
 and six stack SM constructions with just
 the SM at low energies is performed in \cite{kokosneo}.
 It appears \cite{kokosneo} that there is a
 special class of D5 vacua in four, five and six stacks
 of SM embeddings that have the same low energy
 effective theory suggesting that these theories are
 connected at the infrared.

 The purpose of this work is 
to
present further three generation four stack 
string models (PS-I class)
that are based on
the PS $G_{422}$ group, and contain at low energy
exactly
the standard model spectrum,
namely $SU(3)_C \t SU(2)_L \t U(1)_Y$,
without any extra chiral fermions and/or extra gauge
group factors.

We will exhibit the systematics of using {\em extra}
branes with non-trivial intersection numbers with
the color and leptonic branes. The use of
these {\em extra} branes
will serve as a novel mechanism
of scalar singlet generation and breaking of the
$U(1)$'s surviving massless the Green-Schwarz
mechanism.
The presence of the {\em extra} brane mechanism will
be applied to both
PS-A and PS-I classes of PS GUT models. 
We should note that these {\em extra} branes are quite different from
the use of hidden branes, used in non-GUT based
 D6-brane model building
examined in \cite{louis2, kokos, kokos2}. In the
latter models the use of
additional D6 branes needed to satisfy the RR tadpole cancellation 
conditions did not charge the SM chiral fermions and
thus these extra
branes could be characterized as hidden one's. In the
present context the use of
{\em extra} D6 branes cannot be characterized as hidden,
as
there are fields charged under the
{\em extra} branes symmetry group.

 The four-dimensional classes of models we study
are non-supersymmetric
intersecting brane
constructions.
The basic structure behind the models includes
D6-branes intersecting each other at non-trivial angles,
in an orientifolded
factorized six-torus,
where O$_6$ orientifold planes are on top of D6-branes.  
\newline
The proposed classes of models have some distinctive features :
\begin{itemize}

\item   The models start, we neglect for the time being
the presence of {\em  extra} branes,
with a
gauge group at the string
scale $U(4) \times
U(2) \times U(2) \times U(1)$. At the scale of symmetry
breaking of the left-right symmetry, $M_{GUT}$, the initial symmetry group
breaks to the the standard model  
$SU(3)_C \times SU(2)_L \times U(1)_Y $
 augmented with an extra anomaly free $U(1)$ symmetry.
 The additional $U(1)$ symmetry breaks by the vev of
 charged singlet
 scalars to the SM itself at a scale set by its vev.
 The singlets responsible for breaking the
 $U(1)$ symmetry are obtained by
 demanding that certain open string sectors of the
 non-SUSY model respect
 $N=1$ supersymmetry.

\item  Neutrinos gets a mass 
   of the right order, consistent with the LSND oscillation 
experiments, 
from a see-saw mechanism of the Frogatt-Nielsen 
type.
The structure of Yukawa couplings involved in the see-saw 
mechanism supports the smallness of neutrino masses thus
generating a hierarchy in consistency with neutrino oscillation
experiments.

\item  Proton is stable due to the fact that baryon number is an 
unbroken gauged 
global symmetry surviving at low energies and no colour triplet
couplings that could mediate
proton decay exist. 
Thus a gauged baryon number provides a natural explanation for 
proton stability.
As in the models of \cite{louis2, kokos1, kokos, kokos2,
crema, kokosneo}
the baryon number associated
$U(1)$ gauge boson becomes massive through its couplings to Green-Schwarz
mechanism. That has an an immediate effect that baryon number is surviving
as a global symmetry to low energies providing for a natural explanation
for proton stability in general brane-world scenarios.

\item 
The model uses small Higgs representations in the adjoint
to break the PS symmetry,
instead of using large
Higgs representations, .e.g. 126 like in the standard $SO(10)$ models.

\item
The bidoublet Higgs fields $h$ responsible for
electroweak symmetry breaking
do not get charged under the global $U(1)$ and thus
lepton number is
not broken at the standard model.

\item {\em Extra} branes, with non-trivial intersection
numbers with
the colour $a$- and the leptonic $d$-brane, in addition
to the
imposition of N=1 SUSY in sectors coming from the
intersections of the hidden with the leptonic branes, are
being used
to engineer the presence of only the SM at low energy.
The {\em extra} branes are added in single pieces, each one
being associated with a single $U(1)$, e.g. in the
presence of
two (2) {\em extra} $U(1)$ branes the number of
extra $U(1)$'s,
which survive massless the Green-Schwarz mechanism, is
three, and the
theory looks in practical terms like a six-stack model.

\end{itemize}

The paper is organized as follows. 
 In section two 
 we describe the general rules for
 building chiral models in orientifolded $T^6$
 compactifications and the possible open string
 sectors. In section 3, we discuss the basic
 fermion and scalar structure of the PS-I class
 of models that will mainly focus in
 this work.
 In section 4, we make a parenthesis in our study and
discuss the role of the {\em extra} branes in the PS-A class of
models of \cite{kokos1}. The methods  described
in creating singlets scalars fields will serve us
as a prototype for an application to the PS-I class
of models. In section 5, we discuss the parametric
solutions to the
RR tadpoles for the PS-I class of models where we will
be
focusing our attention from now on. In section
6 we discuss the cancellation of $U(1)$ anomalies in
the presence of a 
generalized 
Green-Schwarz (GS) mechanism
and {\em extra} $U(1)$ D6 branes. In subsection 7.1
we discuss the conditions for the absence of tachyons
in the models as well describing the PS breaking Higgs and
the electroweak symmetry breaking Higgs fields.
In subsection 7.2 we discuss the presence of N=1
supersymmetric sectors and {\em extra} sector branes.
The presence of N=1 SUSY creates scalar singlets which
are necessary
to make some unwanted fermions massive enough
to disappear
from the low energy spectrum.
In subsection 7.3 we discuss the breaking of the
surviving the Green-Schwarz mechanism massless $U(1)$'s
with the use of singlets coming from the non-trivial
$N=1$ SUSY intersections of the {\em extra} branes and leptonic branes.
In section 8 we examine the problem of
neutrino masses. We also show that all
additional exotic fermions beyond those of SM become
massive
and disappear from the low energy spectrum.
In this section, we describe in detail how the presence
of supersymmetry
in particular sectors of the theory realizes the
particular couplings
taking part in the see-saw mechanism.
Section 9 contains our conclusions. Finally, 
Appendix A, includes the conditions for the
absence of tachyonic modes
in the spectrum of the PS-I class of models.

\section{Model structure and the rules of computing the
spectrum}

Next, we describe the construction of the PS classes of
models. They are based on  
type I string with D9-branes compactified on a six-dimensional
orientifolded torus $T^6$, 
where internal background 
gauge fluxes on the branes are turned on. 
By performing a T-duality transformation on the $x^4$,
$x^6$, $x^8$,
directions, the D9-branes with fluxes are translated
into D6-branes
intersecting at 
angles. The branes are not paralled to the
orientifold planes.
We assume that the D$6_a$-branes are wrapping 1-cycles 
$(n^i_a, m^i_a)$ along each of the $T^2$
torus of the factorized $T^6$ torus, that is we assume
$T^6 = T^2 \times T^2 \times T^2$.

In order to build a general PS model we consider
four stacks of D6-branes giving rise to their
world-volume to an initial
gauge group $U(4)_c \times U(2)_L \times U(2)_R
\times U(1)$ at the string
scale.
In addition, we consider the addition of NS B-flux, 
which makes the tori tilted, 
and leads to the effective tilted wrapping numbers, 
\beq
(n^i, m ={\tilde m}^i + {n^i}/2);\; n,\;{\tilde m}\;\in\;Z,
\label{na2}
\eeq
allowing semi-integer values for the m-numbers.  
\newline
Because of the $\Omega {\cal R}$ symmetry, 
where $\Omega$ is the worldvolume 
parity and $\cal R$ is the reflection on the T-dualized coordinates,
\beq
T (\Omega )T^{-1}= \Omega {\cal R},
\label{dual}
\eeq
 each D$6_a$-brane
1-cycle, must be accompanied by
its $\Omega {\cal R}$ partner $(n^i_a, -m^i_a)$.

Chiral fermions are obtained by stretched open strings
between
intersecting D6-branes \cite{bele}. 
The chiral spectrum of the models is obtained after
solving simultaneously
the intersection 
constraints coming from the existence of the different
sectors together
with the RR
tadpole cancellation conditions.

There are a number of different sectors, 
which should be taken into account when computing the chiral spectrum.
We denote the action of 
$\Omega R$ on a sector $\a, \b$, by ${\a}^{\star}, {\b}^{\star}$,
respectively.
The possible sectors are:

\begin{itemize}
 
\item The $\a \b + \b \a$ sector: involves open strings stretching
between the
D$6_{\a}$ and D$6_{\b}$ branes. Under the $\Omega R$ symmetry 
this sector is mapped to its image, ${\a}^{\star} {\b}^{\star}
+ {\b}^{\star} {\a}^{\star}$ sector.
The number, $I_{\a\b}$, of chiral fermions in this sector, 
transforms in the bifundamental representation
$(N_{\a}, {\bar N}_{\a})$ of $U(N_{\a}) \times U(N_{\b})$, and reads
\beq
I_{\a\b} = ( n_{\a}^1 m_{\b}^1 - m_{\a}^1 n_{\b}^1)( n_{\a}^2 m_{\b}^2 -
m_{\a}^2 n_{\b}^2 )
(n_{\a}^3 m_{\b}^3 - m_{\a}^3 n_{\b}^3),
\label{ena3}
\eeq
 where $I_{\a\b}$
is the intersection number of the wrapped cycles. Note that the sign of
 $I_{\a\b}$ denotes the chirality of the fermion and with $I_{\a\b} > 0$
we denote left handed fermions.
Negative multiplicity denotes opposite chirality.

\item The $\a\a$ sector : it involves open strings stretching on a single 
stack of 
D$6_{\a}$ branes.  Under the $\Omega R$ symmetry 
this sector is mapped to its image ${\a}^{\star}{\a}^{\star}$.
 This sector contain ${\cal N}=4$ super Yang-Mills and if it exists
SO(N), SP(N) groups appear. 
This sector is of no importance to us as we are 
interested in unitary groups.

\item The ${\a} {\b}^{\star} + {\b}^{\star} {\a}$ sector :
It involves chiral fermions transforming into the $(N_{\a}, N_{\b})$
representation with multiplicity given by
\beq
I_{{\a}{\b}^{\star}} = -( n_{\a}^1 m_{\b}^1 + m_{\a}^1 n_{\b}^1)
( n_{\a}^2 m_{\b}^2 + m_{\a}^2 n_{\b}^2 )
(n_{\a}^3 m_{\b}^3 + m_{\a}^3 n_{\b}^3).
\label{ena31}
\eeq
Under the $\Omega R$ symmetry transforms to itself.

\item the ${\a} {\a}^{\star}$ sector : under the $\Omega R$ symmetry is 
transformed to itself. From this sector the invariant intersections
will give 8$m_{\a}^1 m_{\a}^2 m_{\a}^3$ fermions in the
antisymmetric representation
and the non-invariant intersections that come in pairs provide us with
4$ m_{\a}^1 m_{\a}^2 m_{\a}^3 (n_{\a}^1 n_{\a}^2 n_{\a}^3 -1)$ additional 
fermions in the symmetric and 
antisymmetric representation of the $U(N_{\a})$ gauge group.

\end{itemize}

Any vacuum derived from the previous intersection number constraints of the 
chiral spectrum 
is subject to constraints coming from RR tadpole cancellation 
conditions \cite{tessera}. That requires cancellation of 
D6-branes charges \footnote{Taken together with their
orientifold images $(n_a^i, - m_a^i)$  wrapping
on three cycles of homology
class $[\Pi_{\alpha^{\star}}]$.}, wrapping on three cycles with
homology $[\Pi_a]$ and O6-plane 7-form
charges wrapping on 3-cycles with homology $[\Pi_{O_6}]$. In formal terms,
the RR tadpole cancellation  conditions
in terms of cancellations of RR charges in homology, read :
\beq
\sum_a N_a [\Pi_a]+\sum_{\alpha^{\star}} 
N_{\alpha^{\star}}[\Pi_{\alpha^{\star}}] -32
[\Pi_{O_6}]=0.
\label{homology}
\eeq  
Explicitly, the RR tadpole conditions read :
\beqa
\sum_a N_a n_a^1 n_a^2 n_a^3 =\ 16,\nonumber\\
\sum_a N_a m_a^1 m_a^2 n_a^3 =\ 0,\nonumber\\
\sum_a N_a m_a^1 n_a^2 m_a^3 =\ 0,\nonumber\\
\sum_a N_a n_a^1 m_a^2 m_a^3 =\ 0.
\label{na1}
\eeqa
That ensures absence of non-abelian gauge anomalies but not the
reverse.
A comment is in order. It is important to notice
that the RR tadpole cancellation condition can be understood as
a constraint that demands for each gauge group the number of
fundamentals to be equal to the number of bifundamentals.
As a general rule to D-brane model building, by considering $a$ 
stacks of D-brane configurations with 
$N_a, a=1, \cdots, N$, paralled branes, the gauge group appearing is in 
the form $U(N_1) \times U(N_2) \times \cdots \times U(N_a)$. 
Effectively, each $U(N_i)$ factor will give rise to an $SU(N_i)$
charged under the associated $U(1_i)$ gauge group factor that appears in 
the decomposition $SU(N_a) \times U(1_a)$.
A brane configuration with the unique minimal PS particle content
such that intersection numbers, tadpole conditions and various phenomenological
requirements including the absence of exotic representations are accommodated, 
 can be 
obtained by considering initially four stacks of D6-branes yielding an initial
$U(4)_a \times U(2)_b \times U(2)_c \times U(1)_d $ gauge group equivalent
to an $SU(4)_a \times SU(2)_b \times SU(2)_b \times U(1)_a \times U(1)_b 
\times U(1)_c \times U(1)_d $. Thus, in the first instance, 
we can identify, without loss of 
generality, $SU(4)_a$ as the $SU(4)_c$ colour group 
that its breaking could induce 
the usual $SU(3)$ colour group of strong interactions, the $SU(2)_b$ with 
$SU(2)_L$ of weak interactions and $SU(2)_c$ with $SU(2)_R$. 
Note that the condition to satisfy the RR tadpole cancellation
conditions will force us to add the presence of {\em extra}
branes.

\section{The basic fermion  structure}

The basic PS-I class of models that we will center our attention
in this work, will be a three family non-supersymmetric
GUT model with the left-right
symmetric Pati-Salam model structure
$SU(4)_C \times SU(2)_L \times SU(2)_R$.
The open string background on which the models will
be build will be intersesting D6-branes wrapping
on 3-cycles of
decomposable toroidal ($T^6$) orientifolds of type IIA 
in four dimensions.
 
The three generations of quark and lepton 
fields are accommodated into the following
representations :
\beqa
F_L &=& (4, 2, 1) =\ q(3, 2, \frac{1}{6})
+\ l(1, 2, -\frac{1}{2})
\equiv\ (u,\ d,\ l), \nonumber\\
{\bar F}_R &=& ({\bar 4}, 1,  2) =\ {u}^c({\bar 3}, 1, - \frac{2}{3}) +
{d}^c({\bar 3}, 1, \frac{1}{3}) + {e}^c(1, 1, 1) + {N}^c(1, 1, 0) 
\equiv ( {u}^c,  {d}^c,   {l}^c),\nonumber\\ 
\label{na3}
\eeqa
where the quantum numbers on the right hand side of 
(\ref{na3})
are with respect to the decomposition of the $SU(4)_C \times SU(2)_L \times 
SU(2)_R$ under the $SU(3)_C \times SU(2)_L \times U(1)_Y$ gauge group and
${l}=(\nu,e) $ is the standard left handed lepton doublet,
 ${l}^c=({N}^c, e^c)$ are the right handed leptons.
Also the assignment of the accommodation of the quarks and leptons into
the representations $F_L + {\bar F}_R$ is the one appearing in the
spinorial decomposition of the $16$ representation
of $SO(10)$ under the PS gauge group.
\newline
A set of useful fermions appear also in the model
\beqa
\chi_L^1 =\ ( 1,  {\bar 2}, 1),& \chi_R^1 =\ (1, 1, {\bar 2}),\nonumber\\
\chi_L^2 =\ ( 1,  {\bar 2}, 1),& \chi_R^2 =\ (1, 1, {\bar 2})
\label{useful}
\eeqa
These fermions are a general prediction
of left-right symmetric theories as the existence
of these representations
follows from RR tadpole cancellation conditions. 
\newline
The symmetry breaking of the left-right PS symmetry
at the $M_{GUT}$ scale \footnote{In principle
this scale could be as high as the
string scale.}
proceeds through the representations of
the set of Higgs fields,
\beq
H_1 =\ ({\bar 4}, 1, {\bar 2}), \  \
H_2 =\ ( 4, 1, 2),
\label{useful1}
\eeq
 where,
\beqa
H_1 = ({\bar 4}, 1, {\bar 2}) = {u}_H({\bar 3}, 1, \frac{2}{3}) +
{d}_H({\bar 3}, 1, -\frac{1}{3}) + {e}_H(1, 1, -1)+
{\nu}_H(1, 1, 0).
\label{higgs1}
\eeqa
The electroweak symmetry breaking is delivered 
through the bi-doublet Higgs fields $h_i$ $i=1,2$,
fields in the representations 
\beq
h_1 =\ (1, 2, 2),\ \ h_2 =\ (1, {\bar 2}, {\bar 2}) \ .
\label{additi1}
\eeq
Because of the imposition of N=1 SUSY on some open string
sectors, there are also present
 the massless scalar superpartners of the quarks,
 leptons and antiparticles
\beq
{\bar F}_R^H = ({\bar 4}, 1,  2) = {u}^c_H({\bar 3}, 1, -\frac{4}{6})+
{d}^c_H({\bar 3}, 1, \frac{1}{3})+ {e}^c_H(1, 1, 1) + 
{N}^c_H(1, 1, 0) 
\equiv ({u}^c_H, {d}^c_H,  {l}^c_H).\nonumber\\
\label{na368}
\eeq
The latter fields \footnote{
 are replicas
of the fermion fields appearing in the intersection 
$ac$ and receive a vev}
confirm a property shared by all vacua
coming from these type IIA
constructions. That is the replication of
massless fermion spectrum
by an equal number of massive particles in the
same representations and with the same quantum numbers.
\newline
This is the basic fermionic structure appearing in 
the PS models that we
have considered in \cite{kokos1} and will be appearing
later in this work.
Also, a number of charged exotic
fermion fields, which receive a string scale mass, appear
\beq
6(6, 1, 1),\ \ 6({\bar 10}, 1, 1).
\label{beg1}
\eeq

The complete accommodation of the fermion structure of
the PS-I classes of models can be seen
in table one.

\begin{table}[htb] \footnotesize
\renewcommand{\arraystretch}{1.5}
\begin{center}
\begin{tabular}{|c|c||c|c||c||c|c|}
\hline
Fields &Intersection  & $\bullet$ $SU(4)_C \times SU(2)_L \times SU(2)_R$
 $\bullet$&
$Q_a$ & $Q_b$ & $Q_c$ & $Q_d$ \\
\hline
 $F_L$& $I_{ab^{\ast}}=3$ &
$3 \times (4,  2, 1)$ & $1$ & $1$ & $0$ &$0$ \\
 ${\bar F}_R$  &$I_{a c}=-3 $ & $3 \times ({\ov 4}, 1, 2)$ &
$-1$ & $0$ & $1$ & $0$\\
 $\chi_L^1$& $I_{bd} = -6$ &  $6 \times (1, {\ov 2}, 1)$ &
$0$ & $-1$ & $0$ & $1$ \\    
 $\chi_R^1$& $I_{cd} = -6$ &  $6 \times (1, 1, {\ov 2})$ &
$0$ & $0$ & $-1$ &$1$ \\
 $\chi_L^2$& $I_{bd^{\ast}} = -6$ &  $6 \times (1, {\ov 2}, 1)$ &
$0$ & $-1$ & $0$ & $-1$ \\    
 $\chi_R^2$& $I_{cd^{\ast}} = -6$ &  $6 \times (1, 1, {\ov 2})$ &
$0$ & $0$ & $-1$ &$-1$ \\\hline
 $\omega_L$& $I_{aa^{\ast}}$ &  $6 \b^2
 \times (6, 1, 1)$ & $2$ & $0$ & $0$ &$0$ \\
 $z_R$& $I_{aa^{\ast}}$ & $6  \b^2  \times ({\bar 10}, 1, 1)$ &
$-2$ & $0$ & $0$ &$0$  \\
\hline
\end{tabular}
\end{center}
\caption{\small Fermionic spectrum of the $SU(4)_C \times
SU(2)_L \times SU(2)_R$, PS-I class of models together
with $U(1)$ charges. Note that we have not included 
fermions coming from the presence of sectors involving
the {\em extra}
branes.
\label{spectrum8}}
\end{table}

At this point, before we
start discussing the issues of tadpole cancellation for 
the new PS-I classes of models, we will turn our attention
to the PS-A classes of models of \cite{kokos1} as we
want to clarify the role
of the 
presence of the {\em extra} D6 branes in these models.
The same methodology will be applied later in the
present PS-I classes of models.

\section{{\em Extra} brane engineering - the case of PS-A
class of models}

The four stack PS-A class of models of \cite{kokos1} are based on the
same $U(4) \times U(2)_L \times U(2)_L \times U(1)_d$
gauge
structure at the string scale as the PS-I models
that we will be discussing extensively in this
work.
The fermionic field spectrum of PS-A models is the one appearing
in table (\ref{spectrum8001}) and the solution to the RR
tadpole cancellation conditions is given in
table (\ref{ma101}). In the PS-A class of models there are
present the PS breaking Higgs fields $H_1$, $H_2$ from
(\ref{higgs1}) as well the electroweak Higgs fields
(\ref{additi1}).

\begin{table}[htb] \footnotesize
\renewcommand{\arraystretch}{1.5}
\begin{center}
\begin{tabular}{|c|c||c|c||c||c|c|}
\hline
Fields &Intersection  & $\bullet$ $SU(4)_C \times SU(2)_L \times SU(2)_R$
 $\bullet$&
$Q_a$ & $Q_b$ & $Q_c$ & $Q_d$ \\
\hline
 $F_L$& $I_{ab^{\ast}}=3$ &
$3 \times (4,  2, 1)$ & $1$ & $1$ & $0$ &$0$ \\
 ${\bar F}_R$  &$I_{a c}=-3 $ & $3 \times ({\ov 4}, 1, 2)$ &
$-1$ & $0$ & $1$ & $0$\\
 $\chi_L$& $I_{bd} = -12$ &  $12 \times (1, {\ov 2}, 1)$ &
$0$ & $-1$ & $0$ & $1$ \\    
 $\chi_R$& $I_{cd} = -12$ &  $12 \times (1, 1, {\ov 2})$ &
$0$ & $0$ & $-1$ & $-1$ \\\hline
 $\omega_L$& $I_{aa^{\ast}}$ &  $12 \b^2
 {\tilde \epsilon }
 \times (6, 1, 1)$ & $2{\tilde \epsilon }$ & $0$ & $0$ &$0$ \\
 $z_R$& $I_{aa^{\ast}}$ & $6  \b^2 {\tilde \epsilon } \times ({\bar 10}, 1, 1)$ &
$-2{\tilde \epsilon }$ & $0$ & $0$ &$0$  \\
 $s_L$& $I_{dd^{\ast}}$ & $24  \b^2 {\tilde \epsilon }
 \times (1, 1, 1)$ &
$0$ & $0$ & $0$ &$-2{\tilde \epsilon }$  \\
\hline
\end{tabular}
\end{center}
\caption{\small Fermionic spectrum of the $SU(4)_C \times
SU(2)_L \times SU(2)_R$, PS-A class of models together with $U(1)$ charges.
\label{spectrum8001}}
\end{table}

\begin{table}[htb]\footnotesize
\renewcommand{\arraystretch}{2}
\begin{center}
\begin{tabular}{||c||c|c|c||}
\hline
\hline
$N_i$ & $(n_i^1, m_i^1)$ & $(n_i^2, m_i^2)$ & $(n_i^3, m_i^3)$\\
\hline\hline
 $N_a=4$ & $(0, \epsilon)$  &
$(n_a^2, 3 \epsilon \b_2)$ & $({\tilde \epsilon}, {\tilde \epsilon}/2)$  \\
\hline
$N_b=2$  & $(-1, \epsilon m_b^1 )$ & $(1/\beta_2, 0)$ &
$({\tilde \epsilon}, {\tilde \epsilon}/2)$ \\
\hline
$N_c=2$ & $(1, \epsilon m_c^1 )$ &   $(1/\beta_2, 0)$  & 
$({\tilde \epsilon}, -{\tilde \epsilon}/2)$ \\    
\hline
$N_d=1$ & $(0, \epsilon)$ &  $(n_d^2, 6 \epsilon \b_2)$
  & $(-2{\tilde \epsilon}, {\tilde \epsilon})$  \\   
\hline
$\vdots$ & $\vdots$ & $\vdots$  & $\vdots$  \\
\hline
$N_{h}$ & $(1/\beta_1, 0)$ &  $(1/\beta_2, 0)$
  & $(2{\tilde \epsilon}, 0)$  \\\hline
\end{tabular}
\end{center}
\caption{\small
Tadpole solutions for PS-A type models with
D6-branes wrapping numbers giving rise to the 
fermionic spectrum and the SM,
$SU(3)_C \times SU(2)_L \times U(1)_Y$, gauge group at low energies.
The wrappings 
depend on two integer parameters, 
$n_a^2$, $n_d^2$, the NS-background $\beta_i$ and the 
phase parameters $\epsilon = {\tilde \epsilon }= \pm 1$. 
Also there is an additional dependence on the two wrapping
numbers, integer of half integer,
$m_b^1$, $m_c^1$. Note the presence of the
$N_{h}$ {\em extra} $U(1)$ branes.
\label{ma101}}
\end{table}

Also we have defined the angles :
\beqa
\theta_1 \   = \ \frac{1}{\pi} cot^{-1}\frac{ R_1^{(1)}}{ m_b^1 R_2^{(1)}} \ ;\
\theta_2 \  =   \  \frac{1}{\pi} cot^{-1}
\frac{n_a^2 R_1^{(2)}}{3 \beta_2 R_2^{(2)}} \ ;\
\theta_3 \  = \  \frac{1}{\pi} cot^{-1}\frac{2R_1^{(3)}}{R_2^{(3)}},
 \nonumber \\
{\tilde {\theta_2}} \   = 
\ \frac{1}{\pi} cot^{-1}\frac{n_d^2 R_1^{(1)}}{ 6 \b_2 R_2^{(1)}}
\ ;\
{\tilde {\theta_1}} \   = 
\ \frac{1}{\pi} cot^{-1}\frac{ R_1^{(1)}}{ m_c^1 R_2^{(1)}}\ ,\
\label{PSAAA}
\eeqa

The presence of $N=1$ supersymmetry at the sectors
$ac$, $dd^{\star}$ is compatible with the condition
\beq
- {\tilde \vartheta}_1 =  {\tilde \vartheta}_2 =
{\vartheta}_2 = {\vartheta}_3 = \frac{\pi}{4}
\label{conold}
\eeq

Note that in table (\ref{ma101}) we have added
the presence of an arbitrary number of {\em extra} D6 branes
which have a non-zero intersection number with the
colour $a$-brane and the leptonic $d$-brane and thus
additional
massless fermions seem to be produced \footnote{
 The latter issue was not clarified in \cite{kokos1}
 and we will analyze it here.}.
 The {\em extra} $N_h$ branes are necessary to cancel the
 first tadpole condition, the latter giving
\beq
N_h \frac{2{\tilde \epsilon}}{\beta_1 \beta_2} = 16
\label{olstad}
\eeq
Also, the third tadpole condition in (\ref{na1}) gives
the constraint
\beq
2 n_a^2 +\ n_d^2 +\ \frac{1}{\beta_2} (m_b^1 - m_c^1) =\ 0
\label{thirdtad}
\eeq
 The presence of $N_h$ branes correspond to the presence
 of an additional $U(N_h)$ group at the
 string scale. However, for our purposes instead of
 adding an $U(N_h)$ stack of {\em extra} D6 branes to cancel
 the RR tadpoles, we will consider adding $N_h$ $U(1)$
 {\em extra}
 branes positioned at
 $(1/\beta_1, 0) (1/\beta_2, 0) (2, 0)$ each.
 Note that if we had chosen ${\tilde \epsilon} = -1 $, we
 may have added D6 anti-branes at the same position.
That is in order to show that the new massless fields,
appearing from the non-zero intersections 
of each $U(1)$ {\em extra} D6-brane, get a mass, it is enough
to consider only one of the $N_h$ replicas, the latter we
call it $N_{{\hat h}_1}$.
Thus due to the non-zero intersection numbers of the
$N_{{\hat h}_1}$ brane with $a$, $d$ branes we have also
present the
sectors $ah$, $ah^{\star}$, $dh$, $dh^{\star}$. For our
convenience we choose the number of {\em extra} D6-branes to be
\beq
\beta_1 = \beta_2 = \frac{1}{2}, \ \ N_h =\ 2
\label{conve1}
\eeq

Also, we define \footnote{The explanation of the
origin of
the structure of $U(1)$ anomalies will be explained later in
detail.}

\beqa
B_2^3 \wedge [ 2 \epsilon{\tilde \epsilon}]
[(F^b + F^c)],&\nonumber\\
B_2^1 \wedge \left(\epsilon {\tilde \epsilon}[ 
4 n_a^2 \ F^a - 2 n_d^2  F^d +
4 m_b^1 \ F^b + 4 m_c^1 \ F^c]
\right),&\nonumber\\
B_2^o  \wedge \left(3 {\tilde \epsilon} \right) (F^a + F^d).&
\label{anomaPSA}
\eeqa
The extra $U(1)$'s in the presence of the {\em extra} D6-branes
$h^1$, $h^2$ may be defined as follows:
\beq
U(1)^{(4)} =\ (Q^a - Q^d) + (Q^b - Q^c)
+ (F^{h_1} - F^{h_2} ),
\label{proto1}
\eeq
\beq
U(1)^{(5)} =\ (F^{h_1} + F^{h_2} ),
\label{proto2}
\eeq
\beq
U(1)^{(6)} =\ \frac{1}{2}\left(
(Q^a - Q^d) + (Q^b - Q^c)  \right) -F^{h_1} + F^{h_2}. 
\label{proto3}
\eeq
The latter $U(1)$'s are defined in an orthogonal basis.
We don't get additional constraints due to the presence
of the {\em extra} D6-branes in the RR tadpole parameters.
The only issue remaining is the breaking of the
extra $U(1)$'s, (\ref{proto1}), (\ref{proto2}) and (\ref{proto3}).

In the following we will describe the additional sectors
of the
theory due to the presence of the {\em extra} branes, $h^1$,
$h^2$. The analysis concerns the intersections of the
 $h^1$ brane, but it is valid for the $h^2$ brane as well.
 One needs to mimic the procedure for the $h^2$ brane,
 obtaining one more copy of {\em extra} fermions and scalars.
In the analysis below, one needs to set
$\beta_1 = \beta_2 = 1/2$.
Also the net effect of {\em extra} branes is the
creation of enough singlets from the $dh$, $dh^{\star}$
sectors that may be used to break the extra $U(1)$'s
(\ref{proto1}), (\ref{proto2}), (\ref{proto3}).

\begin{itemize}

\item $ah$-{\em sector}

Because $I_{ah} =\frac{3}{\beta_1} > 0$ there are
present $|I_{ah}|
=|\frac{3}{\beta_1}|$ 
fermions $\lambda_1^f$, appearing in
the representations
\beq
 (4, 1, 1 )_{(1, 0, 0, 0; -1)},
\label{dh1}
\eeq
where the fifth-entry is the $h$-brane $U(1)$ charge.

\item $ah^{\star}$-{\em sector}

Since $I_{ah^{\star}} = -\frac{3}{\beta_1} < 0$ there
are present
$|I_{ah^{\star}}|=|\frac{3}{\beta_1}|$ 
fermions ${\tilde \lambda}_2^f$, appearing in
the representations
\beq
 ( {\bar 4}, 1, 1 )_{(-1, 0, 0, 0; -1)}
\label{dh2}
\eeq

\item $dh$-{\em sector}

Since  $I_{dh} = -\frac{12}{\beta_1} < 0$ there are present
$|I_{dh}|=|\frac{12}{\beta_1}|$ fermions ${\tilde \lambda}_3^f $, appearing in
the representations
\beq
 (1, 1, 1 )_{(0, 0, 0, -1; 1)}
\label{dh3}
\eeq

We further require that this sector respects
$N=1$ supersymmetry. The condition for
$N=1$ supersymmetry in this sector is exactly
\beq
-\ \frac{\pi}{2} +\ {\tilde \vartheta}_2
+\ \vartheta_3 = 0
\label{exac1}
\eeq
which is satisfied when ${\tilde \vartheta}_2$,
${\vartheta}_3$ take the value $\pi/4$ in consistency
with (\ref{conold}).
In this case we have also present the
${\tilde \lambda}_3^B$ massless
scalar fields
\beq
(1, 1, 1 )_{(0, 0, 0, -1; 1)}
\label{dh5}
\eeq
The latter scalars receive a vev.
The size of the vev, of order
$M_s$, will
be induced from the size of a coupling
contributing to the mass of
the $\chi_L$ fermions.

\item $dh^{\star}$-{\em sector}

The intersection $I_{dh^{\star}} = -\frac{12}{\beta_1} <
0$, thus there
are present
$|I_{ah^{\star}}|=|\frac{12}{\beta_1}|$ fermions
${\tilde \lambda}_4^f$, appearing in the representations
\beq
( 1, 1, 1 )_{(0, 0, 0, -1; -1)}
\label{dh4}
\eeq

We require that this sector respects
$N=1$ supersymmetry. The condition for $N=1$
supersymmetry in this sector is exactly
the same as in the $dh$ sector, that is (\ref{exac1}).
In this case we have also present $I_{dh^{\star}}$
massless scalar
fields ${\tilde \lambda}_4^B$, appearing
as a linear combination of the representations
\beq
(1, 1, 1 )_{(0, 0, 0, -1; -1)}
\label{dh6}
\eeq
The latter scalars receive a vev with size of order
$M_s$. The latter  will
be induced from the size of a coupling contributing
to the mass of
the $\chi_L$ fermions.

\end{itemize}

We will now show that all fermions receive a mass
and disappear from the low energy spectrum.

\begin{itemize}

\item The mass term for the $\lambda_1^f$ fermion reads:
\beqa
({\bar 4}, 1, 1 )_{(-1, 0, 0, 0; 1)} \
({\bar 4}, 1, 1 )_{(-1, 0, 0, 0; 1)}  \
\langle(4, 1, 2)_{(1, 0, 1, 0, ;0)} \rangle \nonumber\\
\times \langle (4, 1, {\bar 2})_{(1, 0, -1, 0, ;0)} \rangle
\langle (1, 1, 1)_{(0, 0, 0, 1; -1)}\rangle \
\langle (1, 1, 1)_{(0, 0, 0, -1;-1)}\rangle
\label{eksiso1}
\eeqa
or
\beq
{\bar \lambda}_1^f \ {\bar \lambda}_1^f \
\langle H_2 \rangle \
\langle F_R^H \rangle \
\langle \lambda^3_B \rangle \
\langle {\tilde  \lambda }^4_B \rangle
\sim    {\bar \lambda}_1^f \ {\bar \lambda}_1^f \    M_s
\eeq

\item Similarly the mass term for
the  ${\tilde \lambda}_2^f$ fermion reads:

\beqa
(4, 1, 1 )_{(1, 0, 0, 0; 1)} \
(4, 1, 1 )_{(1, 0, 0, 0; 1)} \
\langle ({\bar 4}, 1, 2)_{(-1, 0, 1, 0;0)} \rangle
\nonumber\\
\times \langle ({\bar 4}, 1, {\bar 2})_{(-1, 0, -1, 0; 0)}  \rangle
\langle (1, 1, 1)_{(0, 0, 0, -1; -1)} \rangle \
\langle (1, 1, 1)_{(0, 0, 0, 1; -1)}  \rangle
\label{eksisoaae2}
\eeqa
or
\beq
{\bar {\tilde \lambda}}_2^f \
{\bar  {\tilde \lambda}}_2^f \
\langle {\bar F}_R^H \rangle \
\langle H_1 \rangle \
\langle {\tilde \lambda}^4_B \rangle \
\langle \lambda^3_B \rangle
\sim    {\bar {\tilde \lambda}}_2^f \
{\bar {\tilde  \lambda}}_2^f  \    M_s
\eeq

\item Similarly the mass term for the  ${\tilde \lambda}_3^f$ fermion reads:

\beqa
(1, 1, 1 )_{(0, 0, 0, 1; -1)} \
(1, 1, 1 )_{(0, 0, 0, 1; -1)} \
\langle (1, 1, 1)_{(0, 0, 0, -1; 1)} \rangle  \
\langle (1, 1, 1)_{(0, 0, 0, -1; 1)} \rangle
\label{eksisonai2}
\eeqa
or
\beq
{\bar {\tilde \lambda}}_3^f \ {\bar {\tilde \lambda}}_3^f \
\langle {\tilde \lambda}_3^B \rangle \
\langle {\tilde \lambda}_3^B \rangle  \sim \ M_s \
{\bar {\tilde \lambda}}_3^f \ {\bar {\tilde \lambda}}_3^f
 \eeq

\item Similarly the mass term for the  ${\tilde \lambda}_4^f$ fermion reads:

\beqa
(1, 1, 1 )_{(0, 0, 0, 1; 1)} \
(1, 1, 1 )_{(0, 0, 0, 1; 1)} \
\langle (1, 1, 1)_{(0, 0, 0, -1; -1)} \rangle  \
\langle (1, 1, 1)_{(0, 0, 0, -1; -1)} \rangle
\label{eksisokla2}
\eeqa
or
\beq
{\bar {\tilde \lambda}}_4^f \ {\bar {\tilde \lambda}}_4^f \
\langle {\tilde \lambda}_4^B \rangle \
\langle {\tilde \lambda}_4^B \rangle  \sim \ M_s \
{\bar {\tilde \lambda}}_4^f \ {\bar {\tilde \lambda}}_4^f
 \eeq

\end{itemize}

Thus all fermions, bosons receive vevs, that appear
due to the non-zero intersection numbers of the {\em extra}
$U(1)$ brane with $a$, $d$ branes receive a string
scale mass and disappear from the low energy spectrum.
The surviving massless the Green-Schwarz mechanism
$U(1)$'s (\ref{proto1}), (\ref{proto2}), (\ref{proto3}) may be broken
by vevs of
${\tilde \lambda}_3^B$, ${\tilde \lambda}_4^B$.

A comment is in order. In \cite{kokos1}, the mass of
the left handed weak fermion doublets $\chi_L$ was shown
to receive corrections \footnote{
where we have included the leading contribution of the 
worksheet area connecting the
seven vertices}
 from the coupling
\beq
(1, 2, 1)(1, 2, 1) e^{-A}
\frac{\langle h_2 \rangle \langle h_2 \rangle \langle
{\bar F}_R^H  \rangle \langle H_1 \rangle
\langle {\bar s}_L^H \rangle }{M_s^4}
\stackrel{A \rightarrow 0}{\sim}
\frac{\upsilon^2}{M_s}(1, 2, 1)(1, 2, 1)
\label{co1}
\eeq
or
\beqa
(1, 2, 1)_{(0, 1, 0, -1)} \ (1, 2, 1)_{(0, 1, 0, -1)} 
\langle(1, {\bar 2}, {\bar 2})_{(0, -1, -1, 0)}\rangle \
\langle(1, {\bar 2}, {\bar 2})_{(0, -1, -1, 0)}\rangle &\nonumber\\
\times \ \langle({\bar 4}, 1, 2)_{(-1, 0, 1, 0)}\rangle \
\langle (4, 1, 2)_{(1, 0, 1, 0)}\rangle \ 
\langle {\bf 1}_{(0, 0, 0, 2)}\rangle 
\label{co2}
\eeqa
Thus the mass of $\chi_L$ was of the order
\beq
m_{\ch_L} \sim \frac{\upsilon^2}{M_s}
\eeq
which "localized" $\chi_L$ in the area between
100 -246 GeV. Thus, the necessity to push
the mass of $\chi_L$ over 90 GeV, pushed the
string scale to
be below 650 GeV.

It turns out that the couplings
(\ref{co1}, \ref{co2}) represent only one part
of the couplings that contribute the
lowest order
correction to the $\chi_L$ mass.
The use of the {\em extra} singlets guarantees
the existence of another mass coupling
of the order of the string scale. It involves the {\em extra}
scalars singlets ${\tilde \lambda}_3^B$, $\lambda_4^B$ and reads : 
\beq                                            
(1, 2, 1)(1, 2, 1) e^{-A}
\frac{\langle h_2 \rangle \langle h_2 \rangle \langle
{\bar F}_R^H \rangle \langle H_1 \rangle
\langle {\bar {\tilde \lambda}}_3^B \rangle \langle \lambda_4^B \rangle }
{M_s^5} {\sim}
\frac{\upsilon^2}{M_s} \ e^{-A} \ (1, 2, 1)(1, 2, 1)
\label{kasa1}
\eeq
explicitly, in representation form, given by
\beqa
(1, 2, 1)_{(0, 1, 0, -1)} \ (1, 2, 1)_{(0, 1, 0, -1)} 
\langle(1, {\bar 2}, {\bar 2})_{(0, -1, -1, 0)}\rangle \
\langle(1, {\bar 2}, {\bar 2})_{(0, -1, -1, 0)}\rangle &\nonumber\\
\times \ \langle({\bar 4}, 1, 2)_{(-1, 0, 1, 0)}\rangle \
\langle(4, 1, 2)_{(1, 0, 1, 0)}\rangle \ 
\langle (1, 1, 1 )_{(0, 0, 0, 1; -1)} \rangle  \langle
(1, 1, 1 )_{(0, 0, 0, 1; 1)}\rangle 
\label{kasa111}
\eeqa
where we have included the leading contribution of the 
worksheet area connecting the
eight vertices.
 We should emphasize that the vev of the scalars
 ${\tilde \lambda}_3^B$, ${\tilde \lambda}_4^B$, should be of order $M_s$
 in order for $\chi_L$ to be at least of order
 $\upsilon^2/M_s$. Any other value for
 ${\tilde \lambda}_3^B$, ${\tilde \lambda}_4^B$ sends the mass of $\chi_L$
 at smaller than $M_z$ values.
That is, assuming that the exponential
area factor in front of the coupling is of order one,
beyond the SM fermions, 
there are light fermions surviving between $M_Z$ and the
scale of electroweak symmetry breaking 246 GeV.
The latter fermions appear to be a general
prediction of GUT left-right constructions in the present
D6
intersecting
brane backgrounds.
Exactly the same features
will be found later for the PS-I class of GUT models.

\section{Tadpole cancellation for the PS-I models}

To understand the solution of the RR tadpole cancellation condition, which
will be given in parametric form we 
should make the following comments :
\newline
a) The need to 
realize certain couplings will force us to demand that some 
intersections will
preserve some supersymmetry. Thus some massive
fields will be
``pulled out" from the massive spectrum and become massless.
For example, in order to realize a Majorana mass term for the right
handed neutrinos we will demand that the $ac$ sector
preserves $N=1$ SUSY. That will have as an
immediate effect
to "pull out" from the massive mode spectrum the
${\bar F}_R^H$ particles.
\newline
b)
The intersection numbers, in table (\ref{spectrum8}),  
of the fermions $F_L + {\bar F}_R$ are chosen 
such that $I_{ac} = - 3$, $I_{ab^{\star}} = 3$. Here, $-3$ denotes opposite 
chirality to that of a left handed fermion. 
The choice of additional fermion representations
$(1, {\bar 2} ,1)$,
$(1, 1, {\bar 2})$ is imposed to us by the RR tadpole
cancellation conditions the latter being
equivalent to
$SU(N_a)$ gauge
anomaly cancellation, in this case of $SU(2)_L$,
$SU(2)_R$ gauge anomalies,
\beq
\sum_i I_{ia} N_a = 0,\;\;a = L, R.
\label{ena4}
\eeq


 The theory breaks
just to the standard model $SU(3) \times SU(2) \times U(1)_Y$
at low energies.
 The complete spectrum
of the model appears in table (\ref{spectrum8}). The tadpole solutions
of PS-I
models are presented in table (\ref{spectruma101}).
\newline
c) The mixed anomalies $A_{ij}$ of the four surplus $U(1)$'s 
with the non-abelian gauge groups $SU(N_a)$ of the theory
cancel through a generalized GS mechanism \cite{sa,iru},
involving
 close string modes couplings to worldsheet gauge fields.
 Two combinations of the $U(1)$'s are anomalous and become
 massive through their
 couplings to RR fields, their
 orthogonal 
 non-anomalous combinations survives, combining to a
 single $U(1)$
 that remains massless. Crucial for achieving the RR tadpole
 cancellation is the presence of {\em extra} D6-branes.
 Contrary,
 of what
 is happening in models, with exactly the SM at low energy,
  and a Standard-like structure at the string
  scale \cite{louis2, kokos, kokos2, kokosneo} where
  the {\em extra} branes have no intersection with
  other branes,
  in the GUT models there is a non-vanishing intersection.
  As an immediate consequence, this becomes a singlet generation
  mechanism after imposing $N=1$ SUSY between $U(1)$
   leptonic and the
  $U(1)$ {\em extra} D6-branes.
  Also, contrary to the SM's of
  \cite{louis2, kokos, kokos2, kokosneo}, in the GUT constructions
  on the same open string backgrounds the {\em extra} branes
  do not form a $U(N_h)$ gauge group but rather a
   $U(1)^{N_1} \times U(1)^{N_2} \cdots U(1)^{N_h}$ one.
\newline
d)
The constraint 
\beqa
{\Pi}_{i=1}^3 m^i&=&0.
\label{req1}
\eeqa                  
is not imposed 
and thus
leads to the appearance of the non-trivial chiral
fermion content from the
$aa^{\ast}$, sector with corresponding fermions $\omega_L$,
$z_R$. \newline
e) 
After breaking the PS left-right symmetry at $M_{GUT}$,
the
surviving gauge symmetry
is that of the SM augmented by some anomaly free $U(1)$'s
\footnote{surviving the Green-Schwarz mechanism},
their number depending
on the number of {\em extra} $U(1)$'s that have been added
to satisfy the RR tadpole conditions. In general the
number of the surviving massless the Green-Schwarz mechanism
$U(1)$'s is $1 + n_h$, where $n_h$ is the number of
{\em extra} $U(1)$ branes.

To break the latter $U(1)$ symmetries we will impose
that the $dh$, $dh^{\star}$ sector respects $N=1$ SUSY.
\newline
f) Demanding $I_{ab}=3$, $I_{ac}=-3$,
it implies that the third tori should be tilted.
By looking at the
intersection numbers of table one,  we conclude that
the
b-brane should be paralled to the c-brane and the a-brane should be 
paralled to the d-brane as there is an absence of intersection numbers for 
those branes. Also,
the  
cancellation of the RR crosscap tadpole constraints
is solved from parametric sets of solutions.
They are given in table (\ref{spectruma101}).
We note that we have chosen our {\em extra} D6-branes to be located in 
\beq
(1/\beta_1, 0) \ (1/\beta_2, 0) \ (1, 1/2) 
\label{look}
\eeq

\begin{table}[htb]\footnotesize
\renewcommand{\arraystretch}{2}
\begin{center}
\begin{tabular}{||c||c|c|c||}
\hline
\hline
$N_i$ & $(n_i^1, m_i^1)$ & $(n_i^2, m_i^2)$ & $(n_i^3, m_i^3)$\\
\hline\hline
 $N_a=4$ & $(0, \epsilon)$  &
$(n_a^2, 3 \epsilon \b_2)$ & $({\tilde \epsilon}, {\tilde \epsilon}/2)$  \\
\hline
$N_b=2$  & $(-1, \epsilon m_b^1 )$ & $(1/\beta_2, 0)$ &
$({\tilde \epsilon}, {\tilde \epsilon}/2)$ \\
\hline
$N_c=2$ & $(1, \epsilon m_c^1 )$ &   $(1/\beta_2, 0)$  & 
$({\tilde \epsilon}, -{\tilde \epsilon}/2)$ \\    
\hline
$N_d=1$ & $(0, \epsilon)$ &  $(n_d^2, -6 \epsilon
{\tilde \epsilon}\b_2)$
  & $(2, 0)$  \\   
\hline
$\vdots$ & $\vdots$ & $\vdots$  & $\vdots$  \\
\hline
$N_h$ & $(1/\beta_1, 0)$ &  $(1/\beta_2, 0)$
  & $({\tilde \epsilon}, {\tilde \epsilon}/2)$  \\\hline   
\end{tabular}
\end{center}
\caption{\small
Tadpole solutions for PS-I type models with
D6-branes wrapping numbers giving rise to the 
fermionic spectrum and the SM,
$SU(3)_C \times SU(2)_L \times U(1)_Y$, gauge group at low energies.
The wrappings 
depend on two integer parameters, 
$n_a^2$, $n_d^2$, the NS-background $\beta_i$ and the 
phase parameters $\epsilon = {\tilde \epsilon }= \pm 1$. 
Also there is an additional dependence on the two wrapping
numbers, integer of half integer,
$m_b^1$, $m_c^1$. Note the presence of the
$N_h$ {\em extra} $U(1)$ branes.
\label{spectruma101}}
\end{table}

With the above choice,  all tadpole conditions but
the first and the third
\footnote{We have added an arbitrary number
of $N_D$ $U(1)$ branes which do not contribute to
the rest of the tadpoles and
intersection numbers. This is always an allowed choice.
We chosen not
to exhibit the rest of the tadpoles as they involve the
identity $0=0$.},
 are satisfied, the first giving
\beq
N_h \frac{{\tilde \epsilon}}{\beta_1 \beta_2} = 16.
\label{za}
\eeq
Thus the number of {\em extra} branes depends only on the
NS background in the three tori and the sign of
${\tilde \epsilon}$. We note that when ${\tilde \epsilon}=1$
we add D6 branes, while when ${\tilde \epsilon}=-1$ we add
D6 anti-branes.
For \footnote{In the following examples we choose
${\tilde \epsilon} = 1$.}
 $\beta_1 = \beta_2 = 1$, $N_h = 16$. If
$\beta_1 =1$, $\beta_2 =1/2$ or $\beta_1 = 1/2$, $ \beta_2 =1$,
$N_h = 8$. For 
\beq
\beta_1 = \beta_2 = 1/2, \ N_h =\ 4.
\label{label1co}
\eeq

Also the third tadpole condition gives
\beq
2 n_a^2 +  \frac{1}{\beta_2}(m_b^1 -m_c^1)=0.
\label{ena11}
\eeq
To see clearly the cancellation of tadpoles, we have to choose a
consistent numerical set of wrapping
numbers, e.g.
\beq
\epsilon =\ {\tilde \epsilon} =\ 1,\;n_a^2=-1,
\;m_b^1=2,\;m_c^1= 1,\;
n_d^2=2,\;\b_1=1,
\;\b_2=1/2.
\label{numero1}
\eeq
The latter can be satisfied with the addition of
eight D6-branes
with wrapping numbers $(1, 0)(2, 0)(1, 1/2)$, effectively giving to the 
models
the structure of table (\ref{new}).
\begin{table}[htb]\footnotesize
\renewcommand{\arraystretch}{2}
\begin{center}
\begin{tabular}{||c||c|c|c||}
\hline
\hline
$N_i$ & $(n_i^1, m_i^1)$ & $(n_i^2, m_i^2)$ & $(n_i^3, m_i^3)$\\
\hline\hline
 $N_a=4$ & $(0, 1)$  &
$(-1, 3/2)$ & $(1, 1/2)$  \\
\hline
$N_b=2$  & $(-1, 2 )$ & $(2, 0)$ &
$(1, 1/2)$ \\
\hline
$N_c=2$ & $(1, 1 )$ &   $(2, 0)$  & 
$(1, -1/2)$ \\    
\hline
$N_d=1$ & $(0, 1)$ &  $(2, -3)$
  & $(2, 0)$  \\   
\hline
$N_{h^1} = 1$ & $(1, 0)$ &  $(2, 0)$
  & $(1, 1/2)$  \\
  \hline
$N_{h^2} = 1$ & $(1, 0)$ &  $(2, 0)$
  & $(1, 1/2)$  \\
\hline
$\vdots$ & $\vdots$ & $\vdots$  & $\vdots$  \\
\hline
$N_{h^8}= 1$ & $(1, 0)$ &  $(2, 0)$
  & $(1, 1/2)$  
\\\hline
\end{tabular}
\end{center}
\caption{\small Wrapping number set consistent
with the tadpole constraints (\ref{za}), (\ref{ena11}).
\label{new}}
\end{table}

We note that in the model described by
table (\ref{new}) the non-zero
intersection numbers of the {\em extra} branes with $a$, $d$
branes, will give us just the SM at low energies, in addition to a number of 
$U(1)$'s. The breaking of the latter $U(1)$'s will be facilited by the use
of scalar singlets from the $dh^i$, $i = 1,.., 8$ sectors as we will 
explain later.

g) the hypercharge operator 
 is defined as a linear combination
of the three diagonal generators of the $SU(4)$, $SU(2)_L$, $SU(2)_R$ groups:
\beq
Y = \frac{1}{2}T_{3R}+ \frac{1}{2}T_{B-L},\;T_{3R}=diag(1,-1),\;
T_{B-L}=diag(\frac{1}{3},\frac{1}{3},\frac{1}{3}, -1). 
\label{hyper12}
\eeq 
Also,
\beqa
Q & = & Y   +\ \frac{1}{2} \ T_{3L}.\\
\label{hye1}
\eeqa

\section{Cancellation of U(1) Anomalies}

The mixed anomalies $A_{ij}$ of the $U(1)$'s
with the non-Abelian gauge groups are given by
\beq
A_{ij}= \frac{1}{2}(I_{ij} - I_{i{j^{\star}}})N_i.
\label{ena9}
\eeq
Note that gravitational anomalies cancel since D6-branes never 
intersect O6-planes.
In the orientifolded type I torus models gauge anomaly 
cancellation \cite{iru} proceeds through a 
generalized GS
mechanism \cite{louis2} that makes use of the 10-dimensional RR gauge fields
$C_2$ and $C_6$ and gives at four dimensions
the couplings to gauge fields 
 \beqa
N_a m_a^1 m_a^2 m_a^3 \int_{M_4} B_2^o \wedge F_a &;& n_b^1 n_b^2 n_b^3
 \int_{M_4}
C^o \wedge F_b\wedge F_b,\\
N_a  n^J n^K m^I \int_{M_4}B_2^I\wedge F_a&;&n_b^I m_b^J m_b^K \int_{M_4}
C^I \wedge F_b\wedge F_b\;,
\label{ena66}
\eeqa
where
$C_2\equiv B_2^o$ and $B_2^I \equiv \int_{(T^2)^J \times (T^2)^K} C_6 $
with $I=1, 2, 3$ and $I \neq J \neq  K $. Notice the four dimensional duals
of $B_2^o,\ B_2^I$ :
\beqa
C^o \equiv \int_{(T^2)^1 \times (T^2)^2 \times (T^2)^3} C_6&;C^I \equiv
\int_{(T^2)^I} C_2, 
\label{ena7}
\eeqa
where $dC^o =-{\star} dB_2^o,\; dC^I=-{\star} dB_2^I$.

The triangle anomalies (\ref{ena9}) cancel from the existence of the
string amplitude involved in the GS mechanism \cite{sa} in four 
dimensions \cite{iru}. 
The latter amplitude, where the $U(1)_a$ gauge field couples to one
of the propagating
$B_2$ fields, coupled to dual scalars, that couple in turn to
two $SU(N)$ gauge bosons, is 
proportional \cite{louis2} to
\beq
-N_a  m^1_a m^2_a m^3_a n^1_b n^2_b n^3_b -
N_a \sum_I n^I_a n^J_a n^K_b m^I_a m^J_b m^K_b\; ,
I \neq J, K 
\label{ena8}
\eeq

The RR couplings $B_2^I$ of (\ref{ena66}),
appear into three terms.
In the general case we should have consider
the contribution of $N_h$ $U(1)$ branes. The extra $U(1)$'s
could be shown that are broken by appropriate vevs.
 The latter will be
exhibited for the case of four {\em extra} $U(1)$
branes. This is a minimal choice of {\em extra} branes given the
choices (\ref{look}), (\ref{label1co}).

\beqa
B_2^3 \wedge [ 2 {\tilde \epsilon}]
[-(F^b + F^c)
+ {\tilde \epsilon} F^{{\hat h}_1}
+ {\tilde \epsilon} F^{{\hat h}_2} 
+ {\tilde \epsilon} F^{{\hat h}_3}
+ {\tilde \epsilon} F^{{\hat h}_4}
],&\nonumber\\
B_2^1 \wedge \left(\epsilon {\tilde \epsilon}[ 
4 n_a^2 \ F^a + 2 n_d^2 {\tilde \epsilon} F^d +
4 m_b^1 \ F^b + 4 m_c^1 \ F^c]
\right),&\nonumber\\
B_2^o  \wedge \left(3 {\tilde \epsilon} \right) F^a.&
\label{rr23}
\eeqa

Moreover, analyzing the mixed anomalies 
of the extra $U(1)$'s with the non-abelian gauge groups $SU(4)_c$, 
$SU(2)_R$, $SU(2)_L$ we can see that there are
two anomaly free combinations
$Q_b - Q_c$, $Q_d$.
As can be seen from (\ref{rr23}) two anomalous combinations of $U(1)$'s,
e.g.
 $F^a$, $-(F^b + F^c)
 + {\tilde \epsilon}(F^{{\hat h}_1} + F^{{\hat h}_2} + F^{{\hat h}_3} + 
F^{{\hat h}_4} )$
 become massive through their couplings to RR
 fields $B_2^o$, $B_2^3$. Also another non-anomalous
 combination, which is model dependent, becomes massive
 through its couplings to the RR field $B_2^1$. 
Also there are five non-anomalous $U(1)$'s which are
getting broken by vevs of some scalars, as we
will discuss later. A comment is on order. We reming the reader 
that the presence of four U(1) {\em extra} D6-branes on top of the
four stack
GUT model, makes PS-I class of models to behave
effectively as eight stack models.

They are :
\beqa
U(1)^{(4)} &=& Q^b - Q^c + Q^d + {\tilde \epsilon}
(F^{{\hat h}_1} + F^{{\hat h}_2} - F^{{\hat h}_3} - F^{{\hat h}_4}),  
\nonumber\\
U(1)^{(5)} &=& (Q^b + Q^c ) +
\frac{{\tilde \epsilon}}{2}
( F^{{\hat h}_1} + F^{{\hat h}_2} + F^{{\hat h}_3} + F^{{\hat h}_4}), \nonumber\\
U(1)^{(6)} &=& \frac{1}{3}(Q^b - Q^c + Q_d) +\
\frac{{\tilde \epsilon}}{4}
(- F^{{\hat h}_1} - F^{{\hat h}_2} + F^{{\hat h}_3} + 
F^{{\hat h}_4}),\nonumber\\
U(1)^{(7)} &=& {\tilde \epsilon} (F^{{\hat h}_1} - F^{{\hat h}_2} + 
F^{{\hat h}_3} - F^{{\hat h}_4}),\nonumber\\  
U(1)^{(8)} &=& {\tilde \epsilon} (F^{{\hat h}_1} - F^{{\hat h}_2} - 
F^{{\hat h}_3} + F^{{\hat h}_4})\ .
\label{addu1}
\eeqa
The eight $U(1)$'s are defined in an orthogonal basis.
The orthogonality relations between the latter $U(1)$'s and
the model dependent $U(1)$ field coupled to $B_2^1$
give us the model
dependent constraints
\beq
2 n_a^2 =\ {\tilde \epsilon} n_d^2, 
\label{modede1}
\eeq
\beq
m_b^1 =\ -m_c^1. 
\label{modede2}
\eeq
At this point we should list 
the couplings of the dual scalars $C^I$ of $B_2^I$
required to cancel
the mixed anomalies of the $U(1)$'s with the 
non-abelian gauge groups $SU(N_a)$.
We have included the contribution of the four {\em extra} $U(1)$
branes ${\hat h}_1, \cdots, {\hat h}_4$. 

They are given by :
\beqa
C^o \wedge 2 {\tilde \epsilon} [-(F^b \wedge F^b) + (F^c \wedge F^c)
+ 2{\tilde \epsilon} 
(F^{h_1} \wedge F^{{\hat h}_1} + F^{{\hat h}_2} \wedge
F^{{\hat h}_2} + F^{{\hat h}_3} \wedge
F^{{\hat h}_3}  &&\nonumber\\
+ F^{{\hat h}_4} \wedge
F^{{\hat h}_4})  ], &\nonumber\\
C^3 \wedge  \left(\frac{3 {\tilde \epsilon}}{2}\right) [ (F^a \wedge 
F^a)-4 (F^d \wedge F^d) ],
&\nonumber\\
C^2 \wedge [{\epsilon}{\tilde \epsilon}][
  \frac{n_a^2  }{2}(F^a \wedge
F^a) + m_b^1 (F^b \wedge 
F^b) - m_c^1 (F^c \wedge 
F^c)].&
\label{nonanomal}
\eeqa
As it will be shown later the condition (\ref{modede1})
will be
derived again, when imposing $N=1$ supersymmetry on some
open string sectors.

The choice's of extra $U(1)$'s (\ref{addu1}) is
consistent with electroweak data
in the sense that they do not break lepton number.
That happens because
the 
bidoublet Higgs fields $h_1$, $h_2$ don't get charged.

\section{Higgs sector, $N=1$ SUSY on
intersections and extra U(1)'s}

\subsection{Stability of the configurations and Higgs sector}

We have so far seen the appearance in the R-sector 
of $I_{ab}$ massless fermions
in the D-brane intersections 
transforming under bifundamental representations $N_a, {\bar N}_b$.
 In intersecting 
brane words, besides the actual
presence of massless fermions at each intersection, 
we have evident the presence of an equal number of
 massive bosons, in the NS-sector, in the same representations 
as the massless fermions \cite{luis1}.
Their mass is of order of the string scale and it should be taken 
into account when examining phenomenological applications related to the
renormalization group equations.
However, it is possible that some of those 
massive bosons may become 
tachyonic \footnote{For consequences
when these set of fields may become massless see \cite{cim}.}, 
especially when their mass, that depends on the 
angles between the branes,
is such that is decreases the world volume of the 
3-cycles involved in the recombination process of joining the two
branes into a single one.
Denoting the twist vector by $(\vartheta_1,\vartheta_2,
\vartheta_3,0)$, in the NS open string sector the 
lowest lying states are given by \footnote{
we assume $0\leq\vartheta_i\leq 1$ .}
{\small \beqa
\begin{array}{cc}
{\rm \bf State} \quad & \quad {\bf Mass} \\
(-1+\vartheta_1,\vartheta_2,\vartheta_3,0) & \alpha' M^2 =
\frac 12(-\vartheta_1+\vartheta_2+\vartheta_3) \\
(\vartheta_1,-1+\vartheta_2,\vartheta_3,0) & \alpha' M^2 =
\frac 12(\vartheta_1-\vartheta_2+\vartheta_3) \\
(\vartheta_1,\vartheta_2,-1+\vartheta_3,0) & \alpha' M^2 =
\frac 12(\vartheta_1+\vartheta_2-\vartheta_3) \\
(-1+\vartheta_1,-1+\vartheta_2,-1+\vartheta_3,0) & \alpha' M^2
= 1-\frac 12(\vartheta_1+\vartheta_2+\vartheta_3)
\label{tachdsix}
\end{array}
\eeqa}
Exactly at the point, where one of these masses may become massless we have 
preservation of ${\cal N}=1$ locally. 
The angles at the six different intersections can be expressed
in terms of the parameters of the tadpole solutions.

$\bullet$ {\em Angle structure and Higgs fields for PS-I classes of models}

The angles at the different intersections can be expressed in terms of the 
tadpole solution parameters. 
We define the angles:
\beqa
\theta_1 \   = \ \frac{1}{\pi} cot^{-1}\frac{ R_1^{(1)}}{ \epsilon m_b^1 R_2^{(1)}} \ ;\
\theta_2 \  =   \  \frac{1}{\pi} cot^{-1}
\frac{n_a^2 R_1^{(2)}}{3 \epsilon \beta_2 R_2^{(2)}} \ ;\
\theta_3 \  = \  \frac{1}{\pi} cot^{-1}\frac{2R_1^{(3)}}{R_2^{(3)}},
 \nonumber \\
{\tilde {\theta_2}} \   = 
\ \frac{1}{\pi} cot^{-1}\frac{n_d^2 R_1^{(1)}}{ 6 \epsilon
{\tilde \epsilon} \b_2 R_2^{(1)}}
\ ;\
{\tilde {\theta_1}} \   = 
\ \frac{1}{\pi} cot^{-1}\frac{ R_1^{(1)}}{ m_c^1 R_2^{(1)}}\ ,\
\label{angPSI}
\eeqa
where $R_{i}^{(j)}$, $i={1,2}$ are the compactification radii
for the three $j=1,2,3$ tori, namely
projections 
of the radii 
 onto the cartesian axis $X^{(i)}$ directions when the NS flux B field,
$b^k$, $k=1,2$ is turned on. 

At each of the six non-trivial intersections 
we have the 
presence of four states $t_i , i=1,\cdots, 4$, associated
to the states (\ref{tachdsix}).
 Hence we have a total of
twenty four different scalars in the model.
The setup is seen clearly if we look at figure one.
These scalars are generally massive but for some values of
their angles could become tachyonic (or massless).

Also, if we demand that the scalars associated with (\ref{tachdsix}) and 
PS-I models 
may not be tachyonic,
we obtain a total of eighteen
conditions for the PS-I type models
 with a D6-brane at angles
configuration to be stable.
They are
given in Appendix A.
We don't consider
the scalars from 
the $dh$, $dh^{\star}$ intersections. For these sectors
we will require later that they preserve $N=1$ SUSY. 
As a result all scalars, but one, in these sectors may become
massive.

\begin{figure}
\centering
\epsfxsize=6in
\hspace*{0in}\vspace*{.2in}
\epsffile{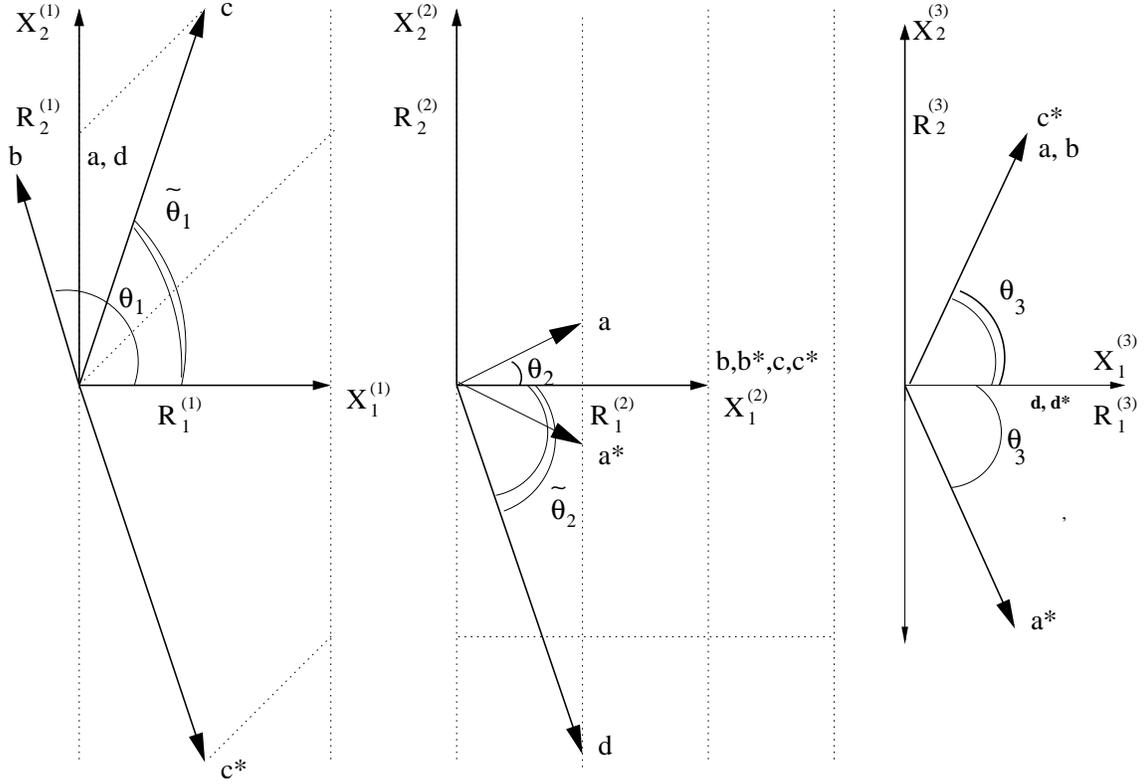}
\caption{\small
Assignment of angles between D6-branes on a
a type I PS-I class of models
based on the initial gauge group $U(4)_C\times {U(2)}_L\times
{U(2)}_R$. The angles between branes are shown on a product of 
$T^2 \times T^2 \times T^2$. We have chosen $ \b_1 =1$, $m_b^1, 
m_c^1, n_a^2 >0$, $\epsilon = {\tilde \epsilon}= 1$. These models
break to low energies to exactly the SM.}
\end{figure}

Lets us now turn our discussion to the Higgs sector of 
PS-I models.
In general there are two different Higgs fields that may
be used to break the PS symmetry.
We remind the reader that they were given in (\ref{useful1}).
The question is if $H_1$, $H_2$ are present in the spectrum of 
PS-I models. 
In general, tachyonic scalars stretching between two 
different branes $\tilde a$, 
$\tilde b$, can be used as Higgs scalars as they can become non-tachyonic
by varying the distance between the branes.
By looking at the $I_{a c^{\star}}$ intersection we can
conclude that the scalar doublets $H^{\pm}$ get
localized.
They come from open strings
stretching
between the $U(4)$ $a$-brane and $U(2)_R$ $c^{\star}$-brane.

\begin{table} [htb] \footnotesize
\renewcommand{\arraystretch}{1}
\begin{center}
\begin{tabular}{||c|c||c|c|c|c||}
\hline
\hline
Intersection & PS breaking Higgs & $Q_a$ & $Q_b$ & $Q_c$ & $Q_d$ \\
\hline\hline
$a c^{\star}$ & $H_1$  &
$1$ & $0$ & $1$ & $0$ \\
\hline
$a c^{\star}$  & $H_2$  & $-1$ & $0$ & $-1$ & $0$  \\
\hline
\hline
\end{tabular}
\end{center}
\caption{\small Higgs fields responsible for the breaking of 
$SU(4) \times SU(2)_R$ 
symmetry of the 
$SU(4)_C \times SU(2)_L \times SU(2)_R$ type I model with D6-branes
intersecting at angles. These Higgs are responsible for giving
masses to the right handed
neutrinos in a single family.
\label{Higgs}}
\end{table}

The $H^{\pm}$'s come from the NS sector and
correspond to the states \footnote{a similar set of
states was used
in \cite{kokos1} to provide the PS-A model
with left-right breaking symmetry scalars $H^{\pm}$.}
{\small \beqa
\begin{array}{cc}
{\rm \bf State} \quad & \quad {\bf Mass^2} \\
(-1+\vartheta_1, \vartheta_2, 0, 0) & \alpha' {\rm (Mass)}^2_{H^{+}} =
  { {Z_3}\over {4\pi ^2}}\ +\ \frac{1}{2}(\vartheta_2 - 
\vartheta_1) \\
(\vartheta_1, -1+ \vartheta_2, 0, 0) & \alpha' {\rm (Mass)}^2_{H^{-}} =
  { {Z_3}\over {4\pi ^2 }}\ +\ \frac{1}{2}(\vartheta_1
  - \vartheta_2) \\
\label{Higgsmasses}
\end{array}
\eeqa}
where $Z_3$ is the distance$^2$ in transverse space along the third torus, 
$\vartheta_1$, $\vartheta_2$ are the (relative)angles 
between the $a$-, $c^{\star}$-branes in the 
first and second complex planes respectively.  
The presence of scalar doublets $H^{\pm}$ can be seen as
coming from the field theory mass matrix

\beq
(H_1^* \ H_2) 
\left(
\bf {M^2}
\right)
\left(
\begin{array}{c}
H_1 \\ H_2^*
\end{array}
\right) + h.c.
\eeq
where
\beqa
{\bf M^2}=M_s^2
\left(
\begin{array}{cc}
Z_3^{(ac^*)}(4 \pi^2)^{-1}&
\frac{1}{2}|\vartheta_1^{(ac^*)}-\vartheta_2^{(ac^*)}|  \\
\frac{1}{2}|\vartheta_1^{(ac^*)}-\vartheta_2^{(ac^*)}| &
Z_3^{(ac^*)}(4 \pi^2)^{-1}\\
\end{array}
\right),
\eeqa
\vspace{1cm}
The fields $H_1$ and $H_2$ are thus defined as
\beq
H^{\pm}={1\over2}(H_1^*\pm H_2) 
\eeq
where their charges are given in table (\ref{Higgs}). 
Hence the effective potential which 
corresponds to the spectrum of the PS symmetry breaking
Higgs scalars is given by
\beqa
V_{Higgs}\ =\ m_H^2 (|H_1|^2+|H_2|^2)\ +\ (m_B^2 H_1H_2\ +\ h.c)
\label{Higgspot}
\eeqa
where
\beqa 
{m_H}^2 \ =\ {{Z_3^{(ac^*)}}\over {4\pi ^2\alpha '}} &;&
m_B^2\ =\ \frac{1}{2\alpha '}|\vartheta_1^{(ac^*)}-
\vartheta_2^{(ac^*)}|
\label{masillas}
\eeqa
The precise values of $m_H^2$, $m_B^2$, for the
PS-I classes of models are :

\beqa
 {m_H}^2 \ \stackrel{PS-I}{=}\ {
 {(\xi_a^{\prime}+\xi_c^{\prime})^2}\over {\alpha '}},&&
m_B^2\ \stackrel{PS-I}{=}\
\frac{1}{2\alpha '}|\frac{1}{2}+ {\tilde\theta}_1 - {\theta}_2|\ ,
\label{value100}
\eeqa
where $\xi_a^{\prime}$($\xi_c^{\prime}$) is the distance between the
orientifold plane
and the $a$, $c^{\star}$ branes and ${\tilde \theta}_1$, 
$ {\theta}_2$ were defined in
(\ref{angPSI}). Thus

\beqa
m_B^2 &\stackrel{PS-I}
{=}&
\frac{1}{2}|
m^2_{{\bar F}_R} (t_2) +\ m^2_{{\bar F}_R}(t_3)
-\ (m^2_{{\bar F}_R} (t_1) +\ m^2_{{\bar F}_R}(t_3)) |\nonumber\\
&=&
\frac{1}{2}|
m^2_{{\bar F}_R} (t_2) +\ m^2_{{\bar F}_R}(t_3)
-\ m^2_{F_L}(t_1) -\
m^2_{F_L}(t_3)|\nonumber\\
&{=}&
\frac{1}{2}|
m^2_{{\chi}_R^1} (t_2) +\ m^2_{{\chi}_R^1}(t_3)
-\ (m^2_{{\bar F}_R} (t_1) +\ m^2_{{\bar F}_R}(t_3)) |\nonumber\\
&=&
\frac{1}{2}|
m^2_{{\chi}_R^1} (t_2) +\ m^2_{{\chi}_R^1}(t_3)
-\ m^2_{F_L}(t_1) -\
m^2_{F_L}(t_3)|\nonumber\\
&{=}&
1 - \frac{1}{2}|
m^2_{{\chi}_R^2} (t_2) +\ m^2_{{\chi}_R^2}(t_3)
-\ (m^2_{{\bar F}_R} (t_1) +\ m^2_{{\bar F}_R}(t_3)) |\nonumber\\
&=&
1- \frac{1}{2}|
m^2_{{\chi}_R^2} (t_2) +\ m^2_{{\chi}_R^2}(t_3)
-\ m^2_{F_L}(t_1) -\
m^2_{F_L}(t_3)|\nonumber\\
\label{kainour}
\eeqa

For PS-I models the number of Higgs present
is equal to the 
the intersection number product between the $a$-, $c^{\star}$- branes
in the
first and second complex planes, 
\beq
n_{H^{\pm}} \stackrel{PS-I}{=}\
|I_{ac^{\star}}|\ =\ 3.
\label{inter1}
\eeq
A comment is in order.   
For PS-I models the number of PS Higgs is three.
That means that we have
three
intersections and to each one we have a Higgs particle
which is a linear
combination of the Higgs $H_1$ and $H_2$.

More Higgs are present. 
In the $bc^{\star}$ intersection we have present some of
the most useful
Higgs fields of the models. They will be used to
give mass to the
quarks and leptons of the model as well breaking
the electroweak symmetry.
They appear in the representations
$(1, 2, 2)$,  $(1, {\bar 2}, {\bar 2})$ and
 have been defined 
  as $h_1$, $h_2$ in (\ref{additi1}).

\begin{table} 
\begin{center}
\begin{tabular}{||c|c||c|c|c|c||}
\hline
\hline
Intersection & Higgs & $Q_a$ & $Q_b$ & $Q_c$ & $Q_d$ \\
\hline\hline
$bc^{\star}$  & $h_1 = (1, 2, 2)$  &
$0$ & $1$ & $1$ & $0$ \\
\hline
$bc^{\star}$ & $h_2 = (1, {\bar 2}, {\bar 2})$  & $0$ & $-1$ & $-1$ & $0$  \\
\hline
\hline
\end{tabular}
\end{center}                 
\caption{\small Higgs fields present in the intersection $bc^{\star }$
of the
$SU(4)_C \times SU(2)_L \times SU(2)_R$ type I model with D6-branes
intersecting at angles. These Higgs give masses to the quarks and
leptons in a single family and are responsible for 
electroweak symmetry breaking.
\label{Higgs3}}
\end{table}

In the NS sector
the lightest scalar states $h^{\pm}$ originate from
open strings stretching between the
$bc^{\star }$ branes

{\small \beqa
\begin{array}{cc}
{\rm \bf State} \quad & \quad {\bf Mass^2} \\
(-1+\vartheta_1, 0, 0, 0) & \alpha' {\rm (Mass)}^2 =
  { {{\tilde Z}_{23}^{bc^{\star}}}\over {4\pi^2}}\ -\ \frac{1}{2}(\vartheta_1)  \\
 (\vartheta_1, -1, 0, 0) &  \alpha' {\rm (Mass)}^2 =
  { {{\tilde Z}_{23}^{bc^{\star}}}\over {4\pi^2}} +\  \frac{1}{2}(\vartheta_1)
\label{Higgsacstar}
\end{array}
\eeqa}
where ${\tilde Z}_{23}^{bc^{\star}}$ is the relative distance in 
transverse space along the second and
third torus from the orientifold plane, 
$\th_1$, is the (relative)angle 
between the $b$-, $c^{\star}$-branes in the 
first complex plane.

The presence of scalar doublets $h^{\pm}$ can be seen as
coming from the field theory mass matrix

\beq
(h_1^* \ h_2) 
\left(
\bf {M^2}
\right)
\left(
\begin{array}{c}
h_1 \\ h_2^*
\end{array}
\right) + h.c.
\eeq
where
\beqa
{\bf M^2}=M_s^2
\left(
\begin{array}{cc}
Z_{23}^{(bc^*)}(4\pi^2)^{-1}&
\frac{1}{2}|\vartheta_1^{(bc^*)}|  \\
\frac{1}{2}|\vartheta_1^{(bc^*)}| &
Z_{23}^{(bc^*)}(4\pi^2)^{-1}\\
\end{array}
\right),
\eeqa
The fields $h_1$ and $h_2$ are thus defined as
\beq
h^{\pm}={1\over2}(h_1^*\pm h_2) \   \ .
\eeq
The effective potential which 
corresponds to the spectrum of electroweak
Higgs $h_1$, $h_2$ may be written as
\beqa
V_{Higgs}^{bc^{\star}}\ =\ \overline{m}_H^2 (|h_1|^2+|h_2|^2)\ +\ 
(\overline{m}_B^2 h_1 h_2\ +\ h.c)
\label{bcstarpote}
\eeqa
where
\beqa 
\overline{m}_H^2 \ =\ {{{\tilde Z}_{23}^{(bc^*)}}\over {4\pi^2\alpha'}} \ 
& ;&
\overline{m}_B^2\ =\ \frac{1}{2\alpha'}|
\vartheta_1^{(bc^{\star})}|
\label{bchiggs}
\eeqa
The precise values of for PS-I classes of
models $\overline{m}_H^2$, $\overline{m}_B^2$ are
\beqa 
 {\bar m}_{H}^2 \ \stackrel{PS-I}{=}\ { {({\tilde \chi}_b^{(2)}
 +{\tilde \chi}_{c^{\star}}^{(2)})^2 + ({\tilde \xi}_b^{(3)} +
 {\tilde \xi}_{c^{\star}}^{(3)})^2}\over
{\alpha '}}\ ;\
{\bar m}_{B}^2\ \stackrel{PS-I}{=}\ \frac{1}{2\alpha'}
|\tilde\theta_1 + \theta_1 |
\ ;
\label{value1002}
\eeqa
where $\theta_1$,  $\tilde \theta_1$ are defined
in (\ref{theta1}), (\ref{thetatil1}).
Also ${\tilde \chi}_b, {\tilde \chi}_{c^{\star}}$ are the 
distances of the $b$, $c^{\star}$ branes from the
orientifold plane in the second tori and ${\tilde \xi}_b$,
${\tilde \xi}_{c^{\star}}$
are the distances of the $b, c^{\star}$ branes from the orientifold plane
in the third tori. Also, notice that the $b$, $c^{\star}$ branes are paralled along
the second and third tori.
The values of the angles $\theta_1$,  $\tilde \theta_1$ can be expressed
in terms of the scalar masses in the various
intersections. They are given by
\beqa
\frac{1}{\pi}\theta_1 &=& \frac{1}{2} + \frac{1}{2}
( m^2_{{\chi}_R^1} (t_2) +\ m^2_{{\chi}_R^1}(t_3) ) \nonumber\\
&=&  \frac{1}{2}( m^2_{{\chi}_R^2} (t_2) +\ m^2_{{\chi}_R^2}(t_3) ) - 
\frac{1}{2}
\nonumber\\
&=& \frac{3}{2} - \frac{1}{2}(m^2_{F_L} (t_2) +\ m^2_{F_L}(t_3))
\label{theta1}
\eeqa
\beqa
\frac{1}{\pi}{\tilde \theta}_1 &=&  \frac{1}{2}
( m^2_{{\chi}_R^1} (t_2) +\ m^2_{{\chi}_R^1}(t_3) ) - \frac{1}{2} \nonumber\\
&=& \frac{1}{2}( m^2_{{\tilde F}_R^2} (t_2) +\ m^2_{{\tilde F}_R^2}(t_3) ) 
- \frac{1}{2}
\nonumber\\
&=& \frac{1}{2} - \frac{1}{2}(m^2_{{\chi}_R^2 } (t_2) +\ m^2_{{\chi}_R^2 }(t_3))
\label{thetatil1}
\eeqa

The number of $h_1$, $h_2$ fields in the $bc^{\star}$
intersection is given by the intersection number of the 
$b$, $c^{\star}$ branes in the first tori
\beq
n_{h^{\pm}}^{b c^{\star}}\
\stackrel{PS-I}{=}\ |\epsilon(m_c^1 - m_b^1)|.
\label{nadou1}
\eeq
A comment is in order. Because the number of the
electroweak
bidoublets in the PS-I models depends on the difference
$|m_b^1-m_c^1|$, by using (\ref{ena11}), (\ref{modede1}),
(\ref{modede2}),
we get
\beq
m_b^1 - m_c^1 =\ 2 m_b^1 =\ - 2 \b_2 n_a^2.
\label{remnain}
\eeq

Hence, e.g. by choosing $n_a^2= 5,\;\b_2 =1/2$,
$m_b^1 = -5/2$,
we get the constraint
\beq
n_{h^{\pm}}^{b c^{\star}}\stackrel{PS-I}{=} \ 5.
\label{natural}
\eeq 
That is we have effectively choose five
electroweak Higgs bidoublet present, each one appearing
in each intersection as a linear
combination of the $h_1$, $h_2$ fields. 
In this case, a consistent numerical set of wrappings
will be, $\epsilon = {\tilde \epsilon}=1$, $m_c^1=5/2$,
$n_d^2=10$.

\begin{table}[htb]\footnotesize
\renewcommand{\arraystretch}{2}
\begin{center}
\begin{tabular}{||c||c|c|c||}
\hline
\hline
$N_i$ & $(n_i^1, m_i^1)$ & $(n_i^2, m_i^2)$ & $(n_i^3, m_i^3)$\\
\hline\hline
 $N_a=4$ & $(0, 1)$  &
$(5, 3/2)$ & $(1, 1/2)$  \\
\hline
$N_b=2$  & $(-1, -5/2 )$ & $(2, 0)$ &
$(1, 1/2)$ \\
\hline
$N_c=2$ & $(1, 5/2 )$ &   $(2, 0)$  & 
$(1, -1/2)$ \\    
\hline
$N_d=1$ & $(0, 1)$ &  $(10, -3)$
  & $(2, 0)$  \\   
\hline
$N_{h^1} = 1$ & $(2, 0)$ &  $(2, 0)$
  & $(1, 1/2)$  \\
 \hline
$\vdots$ & $\vdots$ & $\vdots$  & $\vdots$  \\
\hline
$N_{h^4} = 1$ & $(2, 0)$ &  $(2, 0)$
  & $(1, 1/2)$  \\\hline
\end{tabular}
\end{center}
\caption{\small Wrapping number set sonsistent
with the constraints (\ref{ena11}),
(\ref{modede1}), (\ref{modede2}),
(\ref{remnain}).
\label{consist123}}
\end{table}

\subsection{$N=1$ SUSY on Intersections}

In this section, we will demand that
certain open string sectors
respect $N=1$ supersymmetry.
The chiral spectrum  
of PS-I classes of models described in
table (\ref{spectrum8}) is massless at this point.
The supersymmetry conditions, will create singlet scalars
which receive vevs and generate masses for the 
otherwise massless
fermions $\chi_L^1$, $\chi_L^2$, $\chi_R^1$, $\chi_R^2$.

In order that $N=1$ SUSY is preserved at some
intersection
between two branes {\em L}, {\em M} we need to satisfy
$\pm \vartheta_{ab}^1 \pm \vartheta_{ab}^2 \pm \vartheta_{ab}^3$ for
some choice of signs, where
$\vartheta_{\alpha \beta}^i$, $i=1,2,3$
are the relative angles of the
branes {\em L}, {\em M} across the three 2-tori.

We note that a Majorana mass term for right neutrinos
will appear once we impose $N=1$ SUSY on an intersection.
That will have as an effect the appearance of the massless scalar
superpartners
of the ${\bar F}_R$ fermions, the ${\bar F}_R^H$'s,
allowing a dimension 5
Majorana mass term for $\nu_R$,
$F_R F_R {\bar F}_R^H {\bar F}_R^H$ (see section 8).

$\bullet$ {\em PS-I models with N=1 SUSY}

We demand that the sector $ac$ respects $N=1$
supersymmetry.
The condition for $N=1$ SUSY on the $ac$-sector
reads \footnote{We have chosen $m_c^1 < 0$.}:
\beq
\pm (\frac{\pi}{2} +\ {\tilde \vartheta}_1) \ \pm  {\vartheta}_2 \
\pm  2\vartheta_3
\ = 0,
\label{condo1}
\eeq       
This condition can be solved by choosing :
\beq
ac \rightarrow (\frac{\pi}{2} + \ {\tilde \vartheta}_1) \ + \vartheta_2 \
- 2\vartheta_3 \ = 0,
\label{condo10}
\eeq       
and thus may be solved by the choice \footnote{We have set
$U^{(i)}= \frac{R_2^{(i)}}{R_1^{(i)}}$, $i=1, 2,3$}
\beq
-{\tilde \vartheta}_1 \ =
\vartheta_2 \ = \vartheta_3 \ = \frac{\pi}{4},
\label{solv1}
\eeq
effectively giving us
\beq
\epsilon m_c^1 \ U^{(1)}= \
\frac{3 \epsilon \b_2}{n_a^2} U^{(2)} = \  
  \frac{1}{2}U^{(3)} = \ \frac{\pi}{4}.
\label{condo3}
\eeq

By imposing $N=1$ SUSY on an intersection a massless
scalar
partner appears in this sector. Thus in the $ac$-sector
it is the massless scalar superpartner
of the fermion ${\bar F}_R$, namely
the ${\bar F}_R^H$,
which is generated.
An additional feature of, see (\ref{condo3}),
SUSY on intersections is that
the complex structure moduli $U^i$ takes specific values, 
thus reducing the degeneracy of moduli
parameters in the theory.

As in the discussion of section 4, the
presence of supersymmetry in particular sectors involving
the {\em extra} branes creates singlet scalars that 
provide the couplings that make massive
some non-SM fermions.
In the following discussion we consider only
one of the two $N_h$ $U(1)$ branes, e.g. the
$U(1)_{N_{h_1}}$. The discussion may be repeated identically
for the other $U(1)$ branes present, e.g. $U(1)_{N_{h_2}}$,
$U(1)_{N_{h_3}}$, $U(1)_{N_{h_4}}$.

Due to the non-zero intersection numbers
of the
$N_{h_1}$ brane with $a$, $d$ branes the following
sectors are
present : $ah$, $ah^{\star}$, $dh$, $dh^{\star}$.

\begin{itemize}

\item $ah$-{\em sector}

Because $I_{ah} = 0$ there are no
fermions \footnote{Obviously, there are no
massless bosons from this sector.}
present from this sector.

\item $ah^{\star}$-{\em sector}

Because $I_{ah^{\star}} = - \frac{3}{\beta_1} < 0$,
there are present
$|I_{ah^{\star}}|=|\frac{3}{\beta_1}|$
fermions $\phi_2^f$ appearing in the representations
\beq
({\bar  4}, 1, 1 )_{(-1, 0, 0, 0; -1)}
\l{dh20}
\eeq

\item $dh$-{\em sector}

Because $I_{dh} = - \frac{6}{\beta_1} < 0$, there are present $|I_{dh}|
=|\frac{6}{\beta_1}|$
fermions $\phi_3^f$, appearing in the representations
\beq
(1, 1, 1 )_{(0, 0, 0, -1; 1)}
\l{dh30}
\eeq

We further require that this sector respects
$N=1$ supersymmetry. The condition for
$N=1$ supersymmetry in this sector is exactly
\beq
\frac{\pi}{2} -\ {\tilde \vartheta}_2
-\ \vartheta_3 =\ 0
\label{exac11}
\eeq
which is satisfied when ${\tilde \vartheta}_2$,
${\vartheta}_3$ take the value $\pi/4$ in consistency
with (\ref{conold}).
The latter condition ( and (\ref{conold})  ) implies
\beq
2 n_a^2 =\ {\tilde \epsilon} n_d^2.
\label{constra1}
\eeq
which is exactly one of the conditions for the extra
$U(1)$'s $U(1)^{(4)}$, $U(1)^{(5)}$, $U(1)^{(6)}$,
to survive massless the Green-Schwarz mechanism!
That is, choosing an orthogonal basis for the anomaly free
$U(1)$'s, the requirement on them to
survive massless the Green-Schwarz mechanism
is equivalent to the creation of singlet scalars from a
$N=1$ supersymmetric intersection between a $U(1)$
{\em extra} brane and the "leptonic" $U(1)$ d-brane.
Thus the presence of $N = 1$
supersymmetric sectors involving {\em extra}
branes in the non-SUSY PS-I classes of GUT models
is equivalent to the presence of the generalized
Green-Schwarz anomaly cancellation mechanism. 
A set of wrapping numbers consistent with this
constraint can be seen
in (\ref{consist123}).

Also present are the $|I_{dh}|$ massless
scalar fields $\phi_3^B$,
appearing in the representations
\beq
(1, 1, 1 )_{(0, 0, 0, -1; 1)}
\l{dh50}
\eeq
The latter scalars receive a vev
which we assume to be
of order of the string scale.
The size of the vev will be induced once we examine
the mass couplings of the $\chi_L^1$ fermion (see also
comments on concluding section).

\item $dh^{\star}$-{\em sector}

Because $I_{dh^{\star}} =  \frac{6}{\beta_1} > 0$,
there are present
$|I_{ah^{\star}}|=|\frac{6}{\beta_1}|$
fermions $\kappa_4^f$,
appearing in the representations
\beq
 ( 1, 1, 1 )_{(0, 0, 0, 1; 1)} \ .
\l{dh40}
\eeq

We require that this sector respects
$N=1$ supersymmetry. The condition for $N=1$
supersymmetry in this sector is exactly
the same as in the $dh$ sector, that is (\ref{exac1}).
In this case we have also present
$|I_{dh^{\star}}|=|\frac{6}{\beta_1}|$ massless scalar
fields $\kappa_4^B$
appearing in the representations
\beq
 (1, 1, 1 )_{(0, 0, 0, 1; 1)} \ .
\l{dh500}
\eeq
The latter scalars receive a vev 
which we assume to be
of order of the string scale. The size of the vev
will be induced once we examine the mass couplings
of $\chi_L^2$ fermions.

\end{itemize}

We will now show that all fermions, appearing from
the non-zero intersections of the {\em extra}
brane $U(1)_{N_{h_1}}$
with the branes $a$, $d$ receive string scale mass.

\begin{itemize}

\item The mass term for the $\phi_2^f$ fermion reads:
\beqa
(4, 1, 1 )_{(1, 0, 0, 0; 1)} \
(4, 1, 1 )_{(1, 0, 0, 0; 1)} \
\langle ({\bar 4}, 1, 2)_{(-1, 0, 1, 0;0)} \rangle\nonumber\\
\times \langle ({\bar 4}, 1, {\bar 2})_{(-1, 0, -1, 0; 0)} \rangle
\langle (1, 1, 1)_{(0, 0, 0, -1; -1)} \rangle \
\langle (1, 1, 1)_{(0, 0, 0, 1; -1)}  \rangle
\label{eksiso2}
\eeqa
or
\beq
{\bar \phi}_2^f \ {\bar \phi}_2^f \
\langle H_1 \rangle \
\langle {\bar F}_R^H \rangle \
\langle {\phi}^4_B \rangle \
\langle \kappa^3_B \rangle
\sim   {\bar \phi}_2^f \ {\bar \phi}_2^f \  M_s
\eeq

\item  The mass term for the $\phi_3^f$ fermion reads:

\beqa
(1, 1, 1 )_{(0, 0, 0, 1; -1)} \
(1, 1, 1 )_{(0, 0, 0, 1; -1)} \
\langle (1, 1, 1)_{(0, 0, 0, -1; 1)} \rangle  \
\langle (1, 1, 1)_{(0, 0, 0, -1; 1)} \rangle
\label{eksisooo2}
\eeqa
or
\beq
{\bar \phi}_3^f \ {\bar \phi}_3^f \
\langle {\phi}_3^B \rangle \
\langle {\phi}_3^B \rangle  \sim \ M_s \
{\bar \phi}_3^f \ {\bar \phi}_3^f
 \eeq

\item The mass term for the $\kappa_4^f$ fermion reads:

\beqa
(1, 1, 1 )_{(0, 0, 0, -1; -1)} \
(1, 1, 1 )_{(0, 0, 0, -1; -1)} \
\langle (1, 1, 1)_{(0, 0, 0, 1; 1)} \rangle  \
\langle (1, 1, 1)_{(0, 0, 0, 1; 1)} \rangle
\label{eksisoas2}
\eeqa
or
\beq
{\bar \kappa}_4^f \ {\bar \kappa}_4^f \
\langle{ \kappa}_4^B \rangle \
\langle {\kappa}_4^B \rangle  \sim \ M_s \ {\bar \kappa}_4^f \ {\bar \kappa}_4^f
 \eeq

\end{itemize}

\subsection{Breaking the extra U(1)'s}

In the standard version of a non-SUSY left-right Pati-Salam
model if the
neutral component of $H_1$ (resp. $H_2$), $\nu_H$,
acquires a vev, e.g. $\langle \nu^H \rangle$, then
the initial gauge symmetry,
$U(4) \times U(2)_L \times U(2)_R
\times U(1)_d$,
  can break to the
standard model gauge group
$SU(3) \t U(2) \t U(1)_Y$ augmented by the extra,
non-anomalous,
$U(1)$. In the PS-I models the initial gauge symmetry is
$U(4) \times U(2)_L \times U(2)_R
\times U(1)_d \times U(1)_{N_{h_1}} \times
U(1)_{N_{h_2}}  \times U(1)_{N_{h_3}} \times U(1)_{N_{h_4}}$. 
After the use of the Green-Schwarz
mechanism and the use of the PS breaking Higgs scalars
the gauge symmetry is
$SU(3) \t U(2) \t U(1)_Y \times U(1)^{(4)} \times
U(1)^{(5)} \times U(1)^{(6)} \times U(1)^{(7)}\times U(1)^{(8)} $. 
By appropriate Higgsing we may break
the extra, beyond the SM, $U(1)$'s.

In the PS-A models, by imposing SUSY on sectors
$dd^{\star}$, $dh$, $dh^{\star}$ we made it possible
to generate
the appearance of the scalar superpartners 
of $s_L$,
${\tilde s }_L$, 
$\phi_3^f$, $\kappa_4^f$.
In this case,
different singlets
${\tilde s }_L$,
are charged under the U(1)'s, $U(1)^{(4)}$,
$U(1)^{(6)}$, and thus break them. Also, 
$\phi_3^B$, $\kappa_4^B$,
get charged under the anomaly free
$U(1)$ symmetries $U(1)^{(5)}$ and thus some of them
may be used to break it. Thus finally,
the PS-A models break to exactly
the much wanted SM gauge group structure,
$SU(3) \otimes SU(2) \otimes U(1)_Y$.

In the case of the non-anomalous
$U(1)$'s (\ref{addu1}) of PS-I class of models
the available singlets that
would break the extra $U(1)$'s
come from the $dh_i$, $dh^{\star}_i$, $i = 1,.,4$ sectors.  
Note that in both PS-A and PS-I classes of Pati-Salam type GUTS 
the extra non-anomalous $U(1)$'s 
have some important phenomenological properties.
In particular they do not charge the PS symmetry
breaking  Higgs
 scalars $H_1$, $H_2$ thus avoiding the appearance of
 axions.

We emphasize that up to this point the only issue remaining
is how we can give non-zero masses to all fermions of
table (\ref{spectrum8}) beyond those of
SM fermions.

\section{Neutrino couplings and fermion masses}

Proton decay is one of the most important problems 
of grand unifies theories. In the standard versions of left-right 
symmetric PS models this problem  
is avoided
as B-L is a gauged symmetry but the problem persists in 
baryon number violating operators of sixth order,
contributing to proton
decay. In the PS-I models proton decay is absent as baryon
number
 survives as a
global symmetry to low energies.
That provides for an explanation for the origin of proton stability
in general brane-world scenarios.
Clearly $Q_a = 3 B + L$ and the baryon B
is given by
\beqa
B = \frac{Q_a + Q_{B-L}}{4}.&  
\label{ba1}
\eeqa

In intersecting brane worlds the usual tree level
SM fermion mass generating trilinear Yukawa couplings
between
the fermion states
$F_L^i$, ${\bar F}_R^j$ and the Higgs fields $H^k$ arise
from the
stretching of the worldsheet between the three
D6-branes which cross
at those intersections.
Its general form for a six dimensional torus
is in the
leading order \cite{luis1},
\beq
Y^{ijk}=e^{- {\tilde A}_{ijk}}, \  {\tilde A}_{ijk} \equiv A_1,
\label{yuk1}
\eeq
where ${\tilde A}_{ijk}$ is the worldsheet area
connecting the three vertices.
The areas of each of the two dimensional torus
involved in
this interaction is typically of order one in string units.
As in \cite{kokos1}, we can without loss of generality
assume
that the areas of the second and third tori are close to
zero.
In this case, the area of the full Yukawa coupling (\ref{yuk1})
takes the form 
\beq
Y^{ijk}= e^{-\frac{R_1 R_2}{a^{\prime}} A_{ijk}},
\label{yuk12}
\eeq
where $R_1$, $R_2$ the radii and 
$ A_{ijk}$ the area of the two dimensional tori in the first complex
plane.  For the dimension five interaction term, like those involved in
the Majorana mass term for the right handed neutrinos
the interaction term is scaled in the form
\beq
Y^{lmni}= e^{- {\tilde A}_{lmni}}, \ {\tilde A}_{lmni} \equiv A_2,
\label{yuk14}
\eeq
where ${\tilde A}_{lmni}$ the worldsheet area connecting the four interaction
vertices. Assuming that the areas of the second and third tori
are close to zero, the four term coupling can be approximated as
\beq
Y^{ijk}= e^{-\frac{R_1 R_2}{a^{\prime}} {\tilde A}_{lmni}},
\label{yuk15}
\eeq
where the area of the ${\tilde A}_{lmni}$ may be of order one in string units.
\newline
Thus the full Yukawa interaction for the chiral spectrum
of the PS-I models
reads :
\beq
\lambda_1 F_L \ {\bar F}_R \ h 
+\ \lambda_2 \frac{F_R {F}_R {\bar F}_R^H {\bar F}_R^H }{M_s},  
\label{era1} 
\eeq
where
\beqa
\lambda_1 \equiv e^{-\frac{R_1 R_2 A_1}{\alpha^{\prime}}},&
\lambda_2 \equiv e^{-\frac{R_1 R_2 A_2}{\alpha^{\prime}}}.
\label{aswq123}
\eeqa
and the Majorana coupling involves the massless
scalar   
superpartners ${\bar F}_R^H$ 
of the antiparticles ${\bar F}_R$.
This coupling is unconventional, in the
sense that the ${\bar F}_R^H$ is generated by
imposing SUSY on a
sector of a non-SUSY model.
 We note the
 presence of $N=1$ SUSY at the sector $ac$.
 As can be seen by comparison
 with (\ref{na368}) 
 the ${\bar F}_R^H$ has a neutral direction that
 receives the vev $\langle H \rangle$.
There is no restriction on the vev of
$F_R^H$ from first
principles and its vev can be anywhere between
the scale of electroweak symmetry breaking and $M_s$.
\newline
The Yukawa term 
\beqa
F_L {\bar F}_R h,&& h=\{h_1, h_2\},
\label{yukbre}
\eeqa
is responsible for the electroweak 
symmetry breaking. This term generates Dirac
masses to up quarks and
neutrinos.
Thus, we 
get
\beq
\lambda_1 F_L {\bar F}_R h  \rightarrow (\lambda_1 \  \upsilon)
(u_i u_j^c + \nu_i N_j^c) + (\lambda_1 \  {\tilde \upsilon}) 
\cdot (d_i d_j^c + e_i e_j^c),  
\label{era2}
\eeq
where we have assumed that 
\beq
\langle h \rangle= \left(
\begin{array}{cc}
\upsilon  & 0 \\
0 & {\bar \upsilon}
\label{era41}
\end{array}
\right)
\label{finalhiggs}
\eeq
We observe that the model gives non-zero tree level
masses
to the fields present. 
These mass relations may be retained at tree level only,
since as the model
has a non-supersymmetric fermion spectrum, it will
receive higher order corrections.  
It is interesting that from (\ref{finalhiggs})
we derive the GUT relation
\cite{ellis}
\beq
m_d =\ m_e \ .
\label{gutscale}
\eeq
as well the unwanted
 \beq
m_u =\ m_{N^c \nu} \ .
\label{gutscale1}
\eeq 

In the case of neutrino masses, 
the  ``unwanted'' (\ref{gutscale1}), associated
to the $\nu - N^c$
mixing,
is modified due to the presence of the Majorana term
in (\ref{era1})
leading to the see-saw mixing type neutrino mass
matrix 
\beqa
\left(
\begin{array}{cc}
\nu&N^c 
\end{array}
\right)\times
 \left(
\begin{array}{cc}
0  & m \\
m & M
\label{era4}
\end{array}
 \right)
\times
\left(
\begin{array}{c}
\nu\\
N^c 
\label{era5}
\end{array}
\right),
\label{er1245}
\eeqa
where
\beq
m= \lambda_1  \upsilon.
\label{eigen1}
\eeq
After diagonalization
the neutrino mass matrix gives us two eigenvalues,
the heavy eigenvalue
\beq
m_{heavy} \approx M =\ \lambda_2 \frac{<H>^2 }{M_s},
\label{neu2}
\eeq
corresponding to the right handed neutrino and 
the light eigenvalue
\beq
m_{light} \approx \frac{m^2}{M} =\ \frac{\lambda_1^2}{\lambda_2  }
\times\frac{\upsilon^2 \ M_s  } { <H>^2} 
\label{neu1}
\eeq
corresponding
to the left handed neutrino \footnote{
The neutrino mass matrix is of
the type of an extended Frogatt-Nielsen
mechanism \cite{fro}
mixing
light with heavy states.}.

Values of the parameters giving us values
for neutrino 
masses between 0.1-10 eV, consistent
with the observed neutrino mixing in neutrino
oscillation measurements, will not be presented here,
as they
have already been discussed in \cite{kokos1}.
The analysis remain the same, as the mass scales as well
the Yukawa coupling parametrization of
the theory do not change. We note that 
the hierarchy of neutrino masses
has been investigated by examining several
different scenaria associated with a light
$\nu_L$ mass including the cases
$ \langle H \rangle = |M_s|$,
$ \langle H \rangle < | M_s |$.
In both cases a hierarchy of neutrino masses in the area
of 0.1-10 V in consistency with neutrino oscillation
experiments can be easily accommodated for a wide choice of
parameters.

Our main focus in this part
is to show that
all additional particles, appearing in table
(\ref{spectrum8}), beyond those of $F_L + {\bar F}_R$,
 get a
heavy mass
and disappear from the low energy spectrum.
The only exception will be the light masses of
$\chi_L^1$, $\chi_L^2$, weak fermion doublets 
which are of order of the electroweak symmetry
breaking scale, e.g. 246 GeV. 
Lets us discuss the latter issue in more detail.
The left handed fermions $\chi_L^1$ receive a mass
from the coupling 
\beq
(1, 2, 1)(1, 2, 1) e^{-A}
 \frac{\langle h_2 \rangle \langle h_2 \rangle
\langle {\bar F}_R^H  \rangle \langle H_1 \rangle
\langle {\phi}_3^B \rangle
\langle \kappa_4^B \rangle}{M_s^5}
\stackrel{A \rightarrow 0}{\sim}
\frac{\upsilon^2}{M_s} \ (1, 2, 1)(1, 2, 1)
\label{ka1sa1}
\eeq
explicitly, in representation form, given by
\beqa
(1, 2, 1)_{(0, 1, 0, -1; 0)} \ (1, 2, 1)_{(0, 1, 0, -1; 0)} 
\langle (1, {\bar 2}, {\bar 2})_{(0, -1, -1, 0; 0)} \rangle \
\langle(1, {\bar 2}, {\bar 2})_{(0, -1, -1, 0; 0)} \rangle &\nonumber\\
\times \ \langle({\bar 4}, 1, 2)_{(-1, 0, 1, 0; 0)}\rangle \
\langle(4, 1, 2)_{(1, 0, 1, 0; 0)}\rangle \ 
\langle (1, 1, 1)_{(0, 0, 0, 1, -1)}\rangle \
\langle (1, 1, 1)_{(0, 0, 0, 1, 1)}\rangle \
\label{ka1sa111}
\eeqa
where we have included the leading contribution of the 
worksheet area connecting the
seven vertices.
Also we have assumed that $\langle {\bar \phi}_3^B \rangle =
\langle \kappa_4^B \rangle =  \langle H_1 \rangle = M_s$.
Any other value for these scalars will lower the mass of
$\chi_L^1$ below $M_z$ something unacceptable.

In the following for simplicity reasons we will set the
leading contribution of the different couplings to
one (e.g. area tends to zero).

Also the left handed fermions $\chi_L^2$
receive an $M_s$  mass
from the coupling 
\beq
(1, 2, 1)(1, 2, 1) 
 \frac{\langle h_2 \rangle \langle h_2 \rangle
\langle {\bar F}_R^H  \rangle \langle H_1 \rangle
\langle {\phi}_3^B \rangle
\langle {\bar \kappa}_4^B \rangle}{M_s^5}
{\sim}
\frac{\upsilon^2}{M_s} \ (1, 2, 1)(1, 2, 1)
\label{ka1sa2}
\eeq
explicitly, in representation form, given by
\beqa
(1, 2, 1)_{(0, 1, 0, -1; 0)} \ (1, 2, 1)_{(0, 1, 0, -1; 0)} 
\langle (1, {\bar 2}, {\bar 2})_{(0, -1, -1, 0; 0)} \rangle \
\langle(1, {\bar 2}, {\bar 2})_{(0, -1, -1, 0; 0)}
\rangle &\nonumber\\
\times \ \langle({\bar 4}, 1, 2)_{(-1, 0, 1, 0; 0)}\rangle \
\langle(4, 1, 2)_{(1, 0, 1, 0; 0)}\rangle \ 
\langle (1, 1, 1)_{(0, 0, 0, -1, 1)}\rangle \
\langle (1, 1, 1)_{(0, 0, 0, -1, -1)}\rangle \
\label{ka1sa112}
\eeqa
Altogether, $\chi_L^1$, $\chi_L^2$,  
 receive a mass of order $\upsilon^2/M_s$ and
 thus are expected to be found between $M_Z$ and
 the scale of electroweak symmetry breaking.

The $\chi_R^1$ doublet fermions receive heavy masses
of order $M_s$ in the following way:
\beq
(1, 1, 2)(1, 1, 2)\frac{\langle H_2 \rangle
\langle F_R^H \rangle \langle {\bar \phi}_3^B \rangle
\langle \kappa_4^B \rangle}{M_s^3}
\label{real21}
\eeq
In explicit representation form
\beqa
 (1, 1, 2)_{(0, 0, 1, -1; 0)}
 \ (1, 1, 2)_{(0, 0, 1, -1; 0)} \
\langle ({\bar 4}, 1, {\bar 2})_{(-1, 0, -1, 0; 0)} \rangle \
\langle ({4}, 1, {\bar 2})_{(1, 0, -1, 0; 0)} \rangle
\nonumber\\
\times \langle (1, 1, 1)_{(0, 0, 0, 1, -1; 0)}\rangle \ 
\langle (1, 1, 1)_{(0, 0, 0, 1, 1; 0)}\rangle 
\label{real200}
 \eeqa
With vevs $<H_2> \sim <F_R^H>  \sim M_s$,
the mass of $\chi_R^1$ is of order $M_s$.

We note that in principle the vevs of
$\phi_3^B$, $\kappa_4^B$
setting the
scale of breaking of the extra anomaly free $U(1)$ could be anywhere 
between $\langle \upsilon \rangle$ and $M_s$.

The $\chi_R^2$ doublet fermions receive heavy masses
of order $M_s$ in the following way:
\beq
(1, 1, 2)(1, 1, 2)\frac{\langle H_2 \rangle
\langle F_R^H \rangle \langle {\phi}_3^B \rangle
\langle {\phi}_3^B \rangle}{M_s^3}
\label{real210}
\eeq
In explicit representation form
\beqa
 (1, 1, 2)_{(0, 0, 1, 1; 0)}  \ (1, 1, 2)_{(0, 0, 1, 1; 0)} \
\langle ({\bar 4}, 1, {\bar 2})_{(-1, 0, -1, 0; 0)} \rangle \
\langle ({4}, 1, {\bar 2})_{(1, 0, -1, 0; 0)} \rangle
\nonumber\\ \times
\langle (1, 1, 1)_{(0, 0, 0, -1, 1; 0)}\rangle \
\langle (1, 1, 1)_{(0, 0, 0, -1, -1; 0)}\rangle 
\label{real20001}
 \eeqa
With vevs $<H_2> \sim <F_R^H>  \sim M_s$,
the mass of $\chi_R^2$ is of order $M_s$.

The 6-plet fermions, $\omega_L$, receive a mass term of
order $M_s$  from the
coupling, 
\beq
({\bar 6}, 1, 1)({\bar 6}, 1, 1)
\frac{\langle H_1  \rangle \langle {F}_R^H
\rangle \langle H_1 \rangle \langle {F}_R^H
\rangle}{M_s^3}
\label{6plet}
\eeq
where we have made use of the $SU(4)$ tensor products
$6 \otimes 6 = 1 + 15 + 20$, $ 4 \otimes 4 = 6 + 10$.
Explicitly, in representation form,
\beqa
({\bar 6}, 1, 1)_{(-2, 0, 0, 0; 0)}
\ ({\bar 6}, 1, 1)_{(-2, 0, 0, 0; 0)}
\langle(4, 1, 2)_{(1, 0, 1, 0; 0)} \rangle \
\langle(4, 1, 2)_{(1, 0, 1, 0; 0)} \rangle &\nonumber\\
\times \ \langle (4, 1, {\bar 2})_{(1, 0, -1, 0; 0)})
\rangle \ \langle(4, 1, {\bar 2})_{(1, 0, -1, 0; 0) })
\rangle
\label{6plet1}
\eeqa

The 10-plet fermions $z_R$ receive
 a heavy mass of order $M_s$ from the coupling
\beq
(10, 1, 1)(10, 1, 1)\frac{\langle {\bar F}_R^H
\rangle \langle {\bar F}_R^H \rangle \langle H_2
\rangle \langle H_2 \rangle}{M_s^3},
\label{10plet}
\eeq
where we have used the
tensor product representations for $SU(4)$,
$10 \otimes 10 = 20 + 35 + 45$,
$20 \otimes {\bar 4} = {\bar 15 } + {\bar 20}$, 
${\bar 20} \otimes {\bar 4} = {\bar 6 } + 10$,
$10 \otimes {\bar 4} =  4  + 36$, $4 \otimes {\bar 4} = 1 + 15$.
Explicitly, in representation form, 

\beqa
(10, 1, 1)_{(2, 0, 0, 0; 0)} (10, 1, 1)_{(2, 0, 0, 0; 0)}
\langle({\bar 4}, 1, 2)_{(-1, 0, 1, 0; 0)}\rangle \
\langle({\bar 4}, 1, {2})_{(-1, 0, 1, 0; 0)}\rangle
&\nonumber\\
\times  \
\langle({\bar 4}, 1, {\bar 2})_{(-1, 0, -1, 0; 0)}\rangle \
\langle({\bar 4}, 1, {\bar 2})_{(-1, 0, -1, 0; 0)}\rangle
\label{10pletagain}
\eeqa
Thus only the chiral fermion content of
the SM fermions remains at low
energy.


\section{Conclusions}

In this work, we have continue our discussion
in \cite{kokos1} of
constructing left-right symmetric $G_{422}$ Pati-Salam
GUT models in the
context
of D6 branes intersecting on compactifications of
type IIA on an
orientifolded factorizable $T^6$ tori.  
The GUT models based on the latter open string backgrounds 
have the unique future of breaking exactly
to the SM at
low energy. They  
are constructed as intersecting number deformations,
around the basic
intersection 
number structure in which the quarks and leptons of
the $G_{422}$ GUT
structure $SU(4)_C \times SU(2)_L \times SU(2)_R$ are
accommodated.

Most important, we presented a new mechanism of generating singlet scalars
in the context of intersecting branes. It amounts in the use
of {\em extra}
 $U(1)$ branes needed in the satisfaction of the RR tadpole
cancellation conditions \footnote{This is to be contrasted 
with models with just the SM at low energy from a Standard -like structure 
at the string scale \cite{louis2, kokos, kokos2, kokosneo},  
where the presence of 
{\em extra} branes has no intersection with the rest of the branes.}, the latter 
having non-trivial intersection numbers with the colour $a$-brane and the 
leptonic $d$-brane. The presence of {\em extra} branes creates singlets scalars 
that may be used to break the additional {\em extra} $U(1)$'s that 
survive massless the Green-Schwarz mechanism. \newline
Equally important, as we showed in subsection 7.2,
the existence
of $N = 1$ supersymmetry conditions in open string sectors
involving the {\em extra} branes \footnote{needed to satisfy
the RR tadpole cancellation conditions},
is equivalent
to the existence of the orthogonality conditions
for the $U(1)$'s surviving massless
the presence of the generalized Green-Schwarz mechanism.

The special form of the solutions to the RR tadpole
cancellation conditions allows exotic, antisymmetric and symmetric,
fermionic representations of the colour degrees of
freedom, arising from
brane-orientifold
image brane, $\alpha \alpha^{\star}$, sectors. 
Interestingly the models have the capacity to accommodate couplings
that give a mass of 
order $M_s$ to all these exotic fermions.

The models have some important phenomenological
features, namely they can easily accommodate
small values of neutrino masses of order 0.1-10 eV in
consistency with neutrino oscillation experiments and a
stable  proton. The stability of the proton is guaranteed
as baryon number is a gauged symmetry 
and survives as
global symmetry to low energies.
Moreover,
colour triplet Higgs couplings that could couple to quarks and leptons
and cause a problem to proton decay are absent in all
classes of models.

Despite the fact, that the non-supersymmetric models we
examined are
free of RR tadpoles and, if the angle
stabilization conditions of Appendix A hold,
free of tachyons,
they will always have closed string NSNS tadpoles 
that cannot all be removed.
Some ways that this might be possible have been suggested in
\cite{blume}, by
freezing
the complex moduli
to discrete values, or by background  redefinition
in terms of 
wrapped metrics \cite{nsns}. However, 
it appears that  a dilaton tadpole will always remain
that could in principle
reintroduce tadpoles in the next leading order.
We note that in NS tadpoles are not existent in supersymmetric
models but the backgrounds that we examined in this work, are 
non-supersymmetric.

Also, we note that the complex structure
moduli \footnote{ As was noted in \cite{kokos1}
the
K\"ahler moduli could be
fixed from its value at the string scale, using relations
involving the product radii (see (\cite{kokos1}) ) but in this way we could 
use a large fine tuning which seems unnatural in a string theory context,
where moduli should be assigned values dynamically.}
can be fixed
to discrete values using the supersymmetry conditions, e.g. see (\ref{condo3}),
and in this way it is possible that some if not all, of the NS tadpoles 
can be removed. We leave this task for a future investigation.

One point that we want to emphasize is that until recently, in
orientifolded
six-torus compactifications   
there was any obvious explanation
for keeping the string scale low \cite{antoba}, e.g. to the 1-100 TeV 
region.
Thus controlling the hierarchy
by making the Planck scale large, while keeping the string scale low,
 by varying the radii of the 
transverse directions \cite{antoba} could not be applied, as there are no
simultaneously transverse torus directions to 
all D6-branes \cite{tessera}. 
However, as was noted in \cite{kokos1} and that is also the case for the 
classes of PS-I models examined in this work, 
there is an alternative mechanism
that keeps the string scale $M_s$ low. In particular the existence
of the light weak doublets $\chi_L^1$, $\chi_L^2$  with mass of order up to
246 GeV, makes a definite prediction for a low string scale 
in the energy range less than 650 GeV.   
That effectively, makes the PS-I class of D6-brane 
models (also the PS-A class) directly 
testable to present or feature accelerators.

We should emphasize that crucial in showing that the
GUT classes of models presented break exactly to the SM at 
low energies was our passive acceptence that there are couplings allowed by 
charge and gauge invariance selection rules that may
give the beyond the SM 
fermions masses, if some scalars get a vev.
However, it should be pointed out that whether of not in the present models
 these scalars get a vev is a highly non-trivial dynamical problem which in
order to be solved precicely, we have to calculate at the string theory level 
the effective potential for these moduli scalars. However, with our present
level of understranding of non-SUSY intersecting
braneworld models this 
is a non-trivial question, as first of all we have to solve the stability 
problem of the configurations, as we have
already commented about. We also note that in the context of intersecting 
branes it is not clear at all, that at the point in the moduli space 
that these 
scalars receive a vev, whether or not the system of recombined branes,
signalling  the presence of tachyon at the minimum of the scalar potential
that would break the gauge symmetry, has a lower energy than the rest of the 
scalars and thus standard electroweak symmetry breaking 
will be preferred \footnote{See also the 2nd reference of 
\cite{luis1} and \cite{sen}
for some discussions relevant to this problem.}.
However, it is absolute amazing that at the 
present level of understanding the intersecting brane worlds, that
we can find models that have all 
the necessary couplings in building classes of models with only the 
Standard model at the low energy limit.

Also, it will be interesting to extend the methods employed in this article,
to GUT groups of the same type in higher stacks \cite{kokos4}.
Summarizing, in the present work, we have shown that
it is possible to consider further four stack
 classes of GUT models with exactly the SM at low energy, the geometry of 
which depends on deforming the basic PS quark-lepton intersection structure
$I_{ab^{\star}}=\ 3$,  $I_{ac}=\ -3$ and the presence
of {\em extra} branes.

\begin{center}
{\bf Acknowledgments}
\end{center}
I am grateful to Luis Ib\'a\~nez,
and Angel Uranga,
for useful discussions. 

\newpage

\section{Tachyon free conditions for classes of PS-I GUTS}

In this appendix we list the conditions, mentioned in
section 5,
  under which the PS-I model D6-brane configurations
of tadpole solutions of table (\ref{spectruma101}),
are tachyon free.
Note that 
the conditions are expressed in terms of the angles
defined in (\ref{angPSI}).

\beqa
\begin{array}{ccccccc}
-(\frac{3\pi}{2} - \vartheta_1) &+& \vartheta_2 &+& 2 {\vartheta}_3 &\geq& 0\\
-(\frac{\pi}{2}+{\tilde \vartheta}_1) &+& \vartheta_2 &+& 
2{\vartheta}_3 &\geq &0 \\
-(-\frac{\pi}{2}+\vartheta_1) &+& {\tilde \vartheta}_2
&+&  {\vartheta}_3 &\geq& 0\\
-(\frac{\pi}{2} + {\tilde \vartheta}_1) &
+&{\tilde \vartheta}_2 &+&
{\vartheta}_3 &\geq& 0 \\
-(\frac{\pi}{2} + { \vartheta}_1) &+&{\tilde \vartheta}_2 &+&
{\vartheta}_3 &\geq& 0 \\
-(\frac{\pi}{2} - {\tilde \vartheta}_1) &+&{\tilde \vartheta}_2 &+&
{\vartheta}_3 &\geq& 0 \\
\\\\
(\frac{3\pi}{2} - \vartheta_1) &-& \vartheta_2 &+& 2 {\vartheta}_3 &\geq& 0\\
(\frac{\pi}{2}+{\tilde \vartheta}_1) &-& \vartheta_2 &+& 
2{\vartheta}_3 &\geq &0 \\
(-\frac{\pi}{2}+\vartheta_1) &-& {\tilde \vartheta}_2
&+&  {\vartheta}_3 &\geq& 0\\
(\frac{\pi}{2} + {\tilde \vartheta}_1) &-&{\tilde \vartheta}_2 &+&
{\vartheta}_3 &\geq& 0 \\
(\frac{\pi}{2} + { \vartheta}_1) &-&{\tilde \vartheta}_2 &+&
{\vartheta}_3 &\geq& 0 \\
(\frac{\pi}{2} - {\tilde \vartheta}_1) &-&{\tilde \vartheta}_2 &+&
{\vartheta}_3 &\geq& 0
\\\\
(\frac{3\pi}{2} - \vartheta_1) &+& \vartheta_2 &-& 2 {\vartheta}_3 &\geq& 0\\
(\frac{\pi}{2}+{\tilde \vartheta}_1) &+& \vartheta_2 &-& 
2{\vartheta}_3 &\geq &0 \\
(-\frac{\pi}{2}+\vartheta_1) &+& {\tilde \vartheta}_2
&-&  {\vartheta}_3 &\geq& 0\\
(\frac{\pi}{2} + {\tilde \vartheta}_1) &
+&{\tilde \vartheta}_2 &-&
{\vartheta}_3 &\geq& 0 \\
(\frac{\pi}{2} + { \vartheta}_1) &+&{\tilde \vartheta}_2 &-&
{\vartheta}_3 &\geq& 0 \\
(\frac{\pi}{2} - {\tilde \vartheta}_1) &+&{\tilde \vartheta}_2 &-&
{\vartheta}_3 &\geq& 0
\label{free}
\end{array}
\eeqa

\end{document}